\documentclass[acmtog]{acmart}
\usepackage{multirow}
\usepackage{graphicx}
\usepackage{color}
\usepackage{amsthm}
\usepackage[normalem]{ulem}
\usepackage[ruled]{algorithm2e}
\SetKwComment{Comment}{/* }{ */}
\usepackage{makecell}

\newcommand{\revision}[1]{{#1}}
\newcommand{\rev}[1]{{#1}}
\newcommand{\revv}[1]{{#1}}

\newcommand{\N}{\mathbb{N}}
\renewcommand{\sout}[1]{}

\AtBeginDocument{%
  \providecommand\BibTeX{{%
    \normalfont B\kern-0.5em{\scshape i\kern-0.25em b}\kern-0.8em\TeX}}}

\setcopyright{acmcopyright}
\acmYear{2024}
\acmDOI{10.1145/3687956}

\acmJournal{TOG}
\acmVolume{43}
\acmNumber{6}
\acmArticle{190}
\acmMonth{12}

\acmSubmissionID{papers\_693s1}

\SetArgSty{textnormal}

\begin{document}

\title{Surface Reconstruction Using Rotation Systems}


\author{Ruiqi Cui}
\affiliation{%
  \institution{Technical University of Denmark}
  \streetaddress{Anker Engelunds Vej 101}
  \city{Lyngby}
  \country{Denmark}}
\email{ruicu@dtu.dk}

\author{Emil Toftegaard Gæde}
\affiliation{%
  \institution{Technical University of Denmark}
  \streetaddress{Anker Engelunds Vej 101}
  \city{Lyngby}
  \country{Denmark}}
\email{etoga@dtu.dk}

\author{Eva Rotenberg}
\affiliation{%
  \institution{Technical University of Denmark}
  \streetaddress{Anker Engelunds Vej 101}
  \city{Lyngby}
  \country{Denmark}}
\email{erot@dtu.dk}

\author{Leif Kobbelt}
\affiliation{%
  \institution{Visual Computing Institute, RWTH Aachen University}
  \streetaddress{1 Th{\o}rv{\"a}ld Circle}
  \city{Aachen}
  \country{Germany}}
\email{kobbelt@cs.rwth-aachen.de}

\author{J. Andreas Bærentzen}
\affiliation{%
  \institution{Technical University of Denmark}
  \streetaddress{Anker Engelunds Vej 101}
  \city{Lyngby}
  \country{Denmark}}
\email{janba@dtu.dk}

\renewcommand{\shortauthors}{R. Cui, E. T. Gæde, E. Rotenberg, L. Kobbelt, and J. A. Bærentzen}

\begin{teaserfigure}
    \centering
    \includegraphics[width=\textwidth]{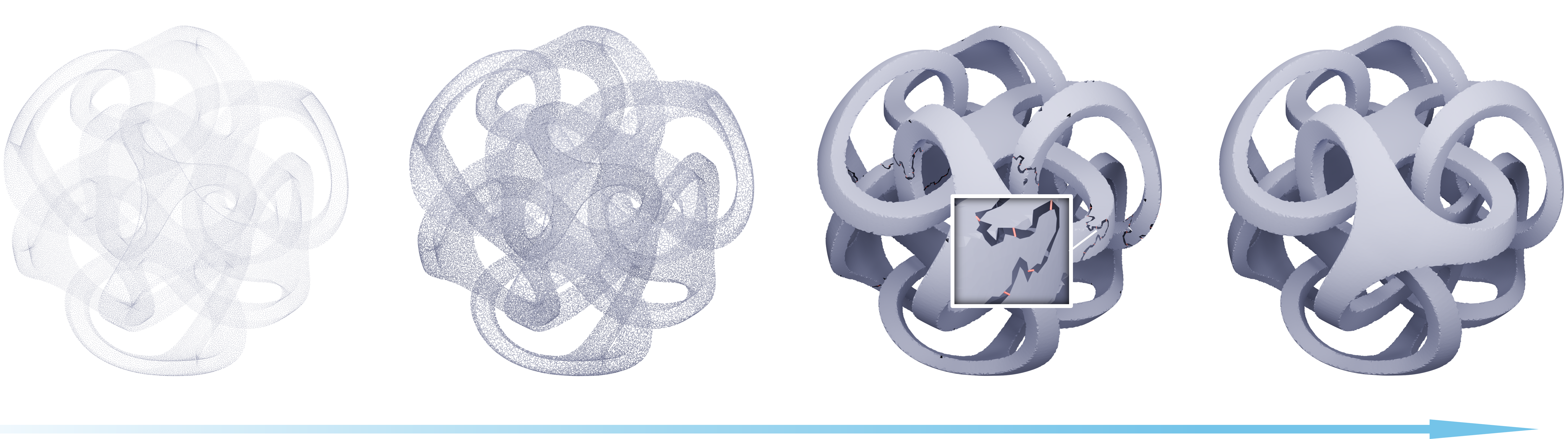}
    \caption{Reconstruction process of a high-genus shape. Starting from the point cloud, a minimum spanning tree (MST) is built. By iteratively inserting proper edges into the MST while keeping the genus to zero, a mesh is provided with multiple cracks. Later, with the red \rev{edges} connecting both sides of the cracks, the genus number of the mesh is increased. Finally, this convoluted high-genus shape is well reconstructed.}
    \label{fig:teaser}
\end{teaserfigure}
\begin{abstract}
Inspired by the seminal result that a graph and an associated rotation system uniquely determine the topology of a closed manifold, we propose a combinatorial method for reconstruction of surfaces from points. Our method constructs a spanning tree and a rotation system. Since the tree is trivially a planar graph, its rotation system determines a genus zero surface with a single face which we proceed to incrementally refine by inserting edges to split faces. In order to raise the genus, special handles are added in a later stage by inserting edges between different faces and thus merging them. We apply our method to a wide range of input point clouds in order to investigate its effectiveness, and we compare our method to several other surface reconstruction methods. It turns out that our approach has two specific benefits over these other methods. First, the output mesh preserves the most information from the input point cloud. Second, our method provides control over the topology of the reconstructed surface. Code is available on \url{https://github.com/cuirq3/RsR}.
\end{abstract}





\maketitle

\nocite{MacLane} 

\section{Introduction}
Reconstruction of surfaces from 3D point clouds is a fundamental problem in geometry processing which continues to attract attention from researchers in the field. Undoubtedly, the main reason for this attention is the great practical utility of reconstruction algorithms due to the abundance of optical devices that can capture 3D point clouds, but it is also important to note that there are many variations of the problem, and the problem is ill-posed. All of these factors contribute to a large number of reconstruction methods with different strengths and weaknesses.

The ill-posed nature of the problem is due to the fact that there is generally no unique solution. Even if we restrict ourselves to only consider methods which connect all points to form a 2-manifold orientable triangle mesh, the number of possible solutions is immense for realistic input sizes. While it is trivial to connect a point to its $k$ nearest neighbors to form a graph, it is far from trivial to form a triangle mesh from this collection of edges, precisely due to the aforementioned combinatorial complexity. However, if we require the output triangle mesh to have disk topology, graph theory provides a solution that to the best of our knowledge has not been considered before. The restriction to disk topology would be problematic, but we show how it is possible to relax this restriction in order to obtain meshes with boundary curves and of arbitrary genus.

Our work is based on the simple observation that a tree (in the graph theoretical sense) is always a \textit{planar} graph, and the planar embedding of a tree is given by a \textit{rotation system} which defines the clockwise ordering of edges incident on each vertex. Thus, if we have a spanning tree connecting all of our points and an associated rotation system, we effectively have a polygonization with a single polygon whose edges are simply the edges of the tree (with each edge of the tree appearing twice). If we imagine that the tree is actually embedded in the 2D plane, the interior of the polygon is simply the entire 2D plane minus the tree.

\begin{figure}[]
  \centering
  \includegraphics[width=0.32\columnwidth]{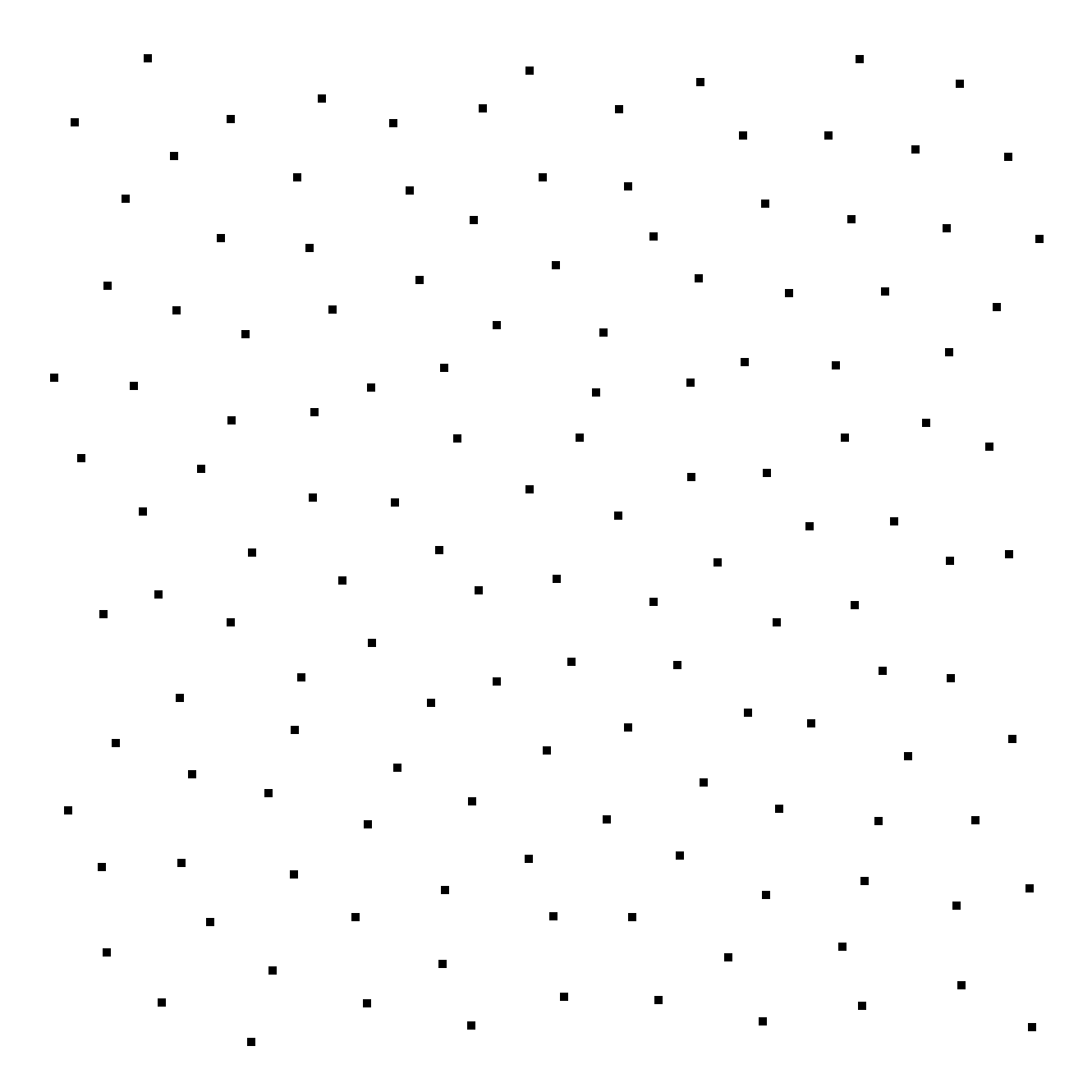}
  \includegraphics[width=0.32\columnwidth]{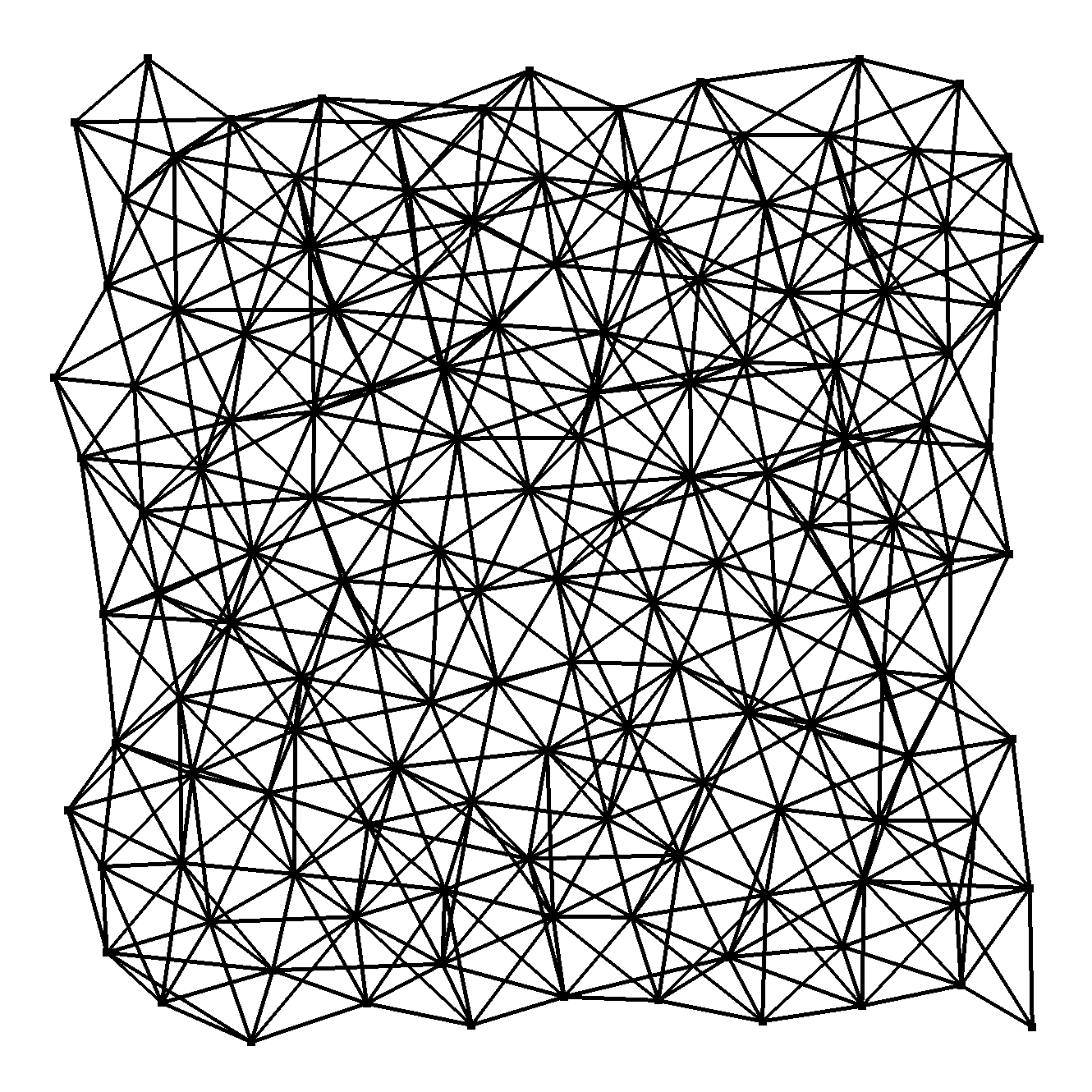}
  \includegraphics[width=0.32\columnwidth]{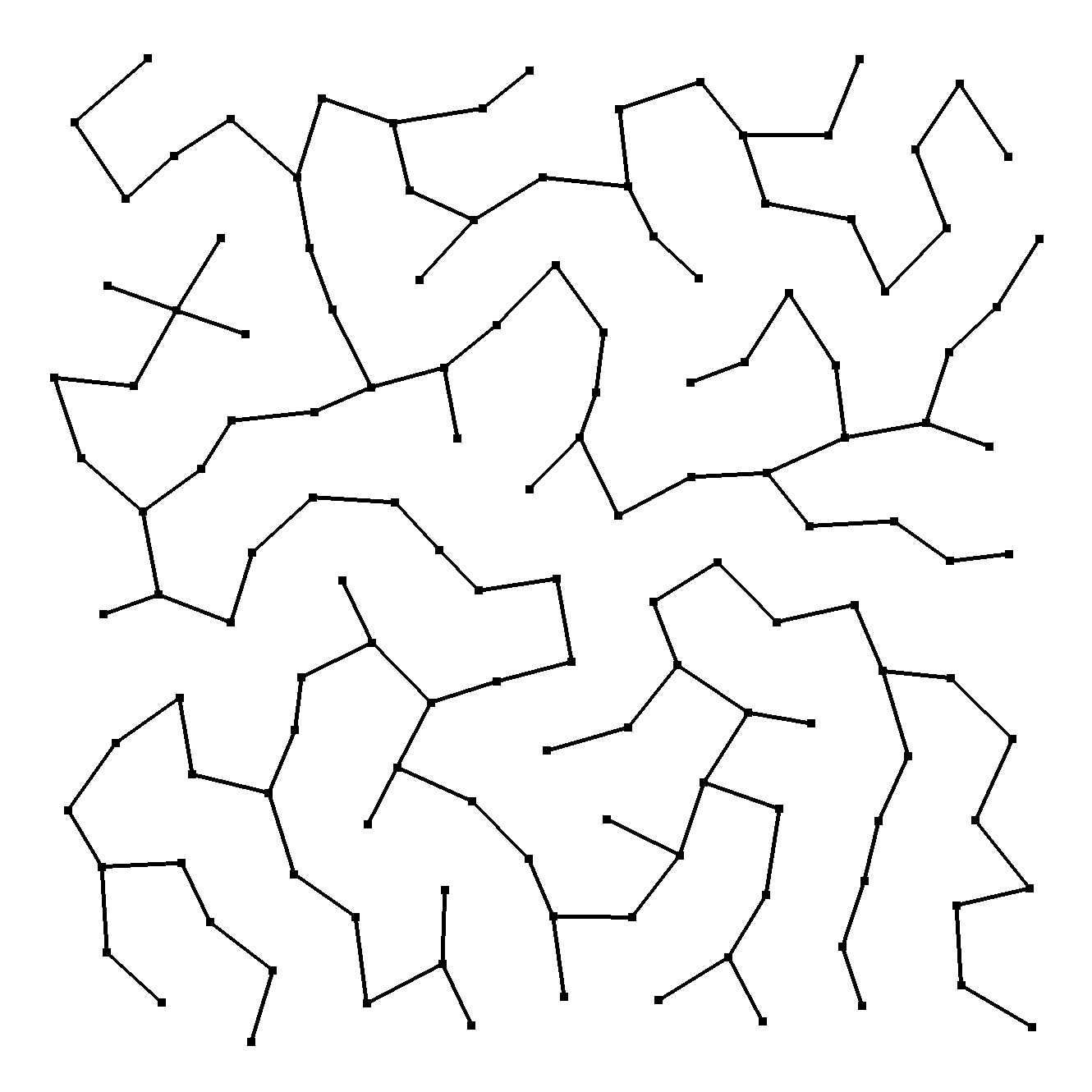}
  \includegraphics[width=0.32\columnwidth]{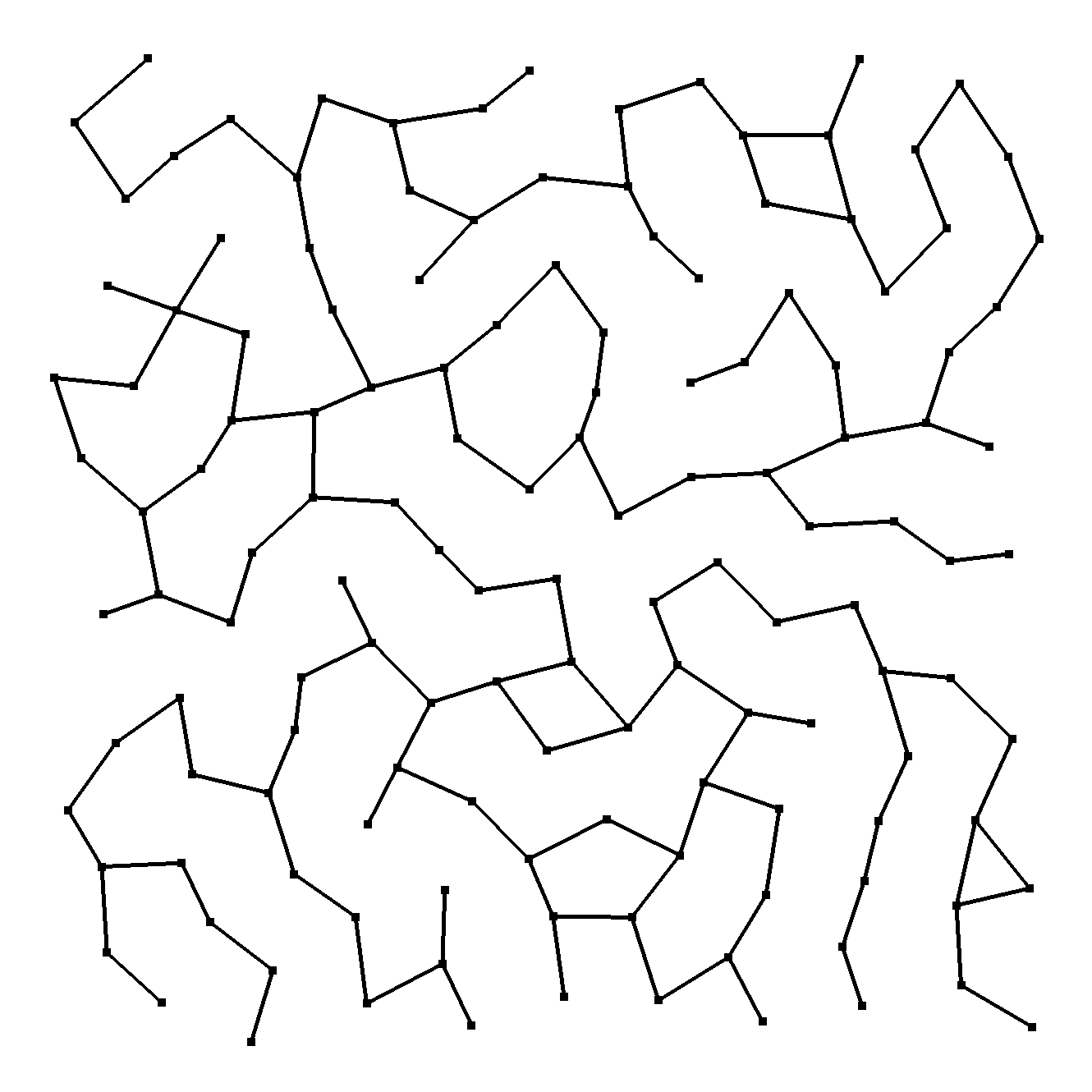}
  \includegraphics[width=0.32\columnwidth]{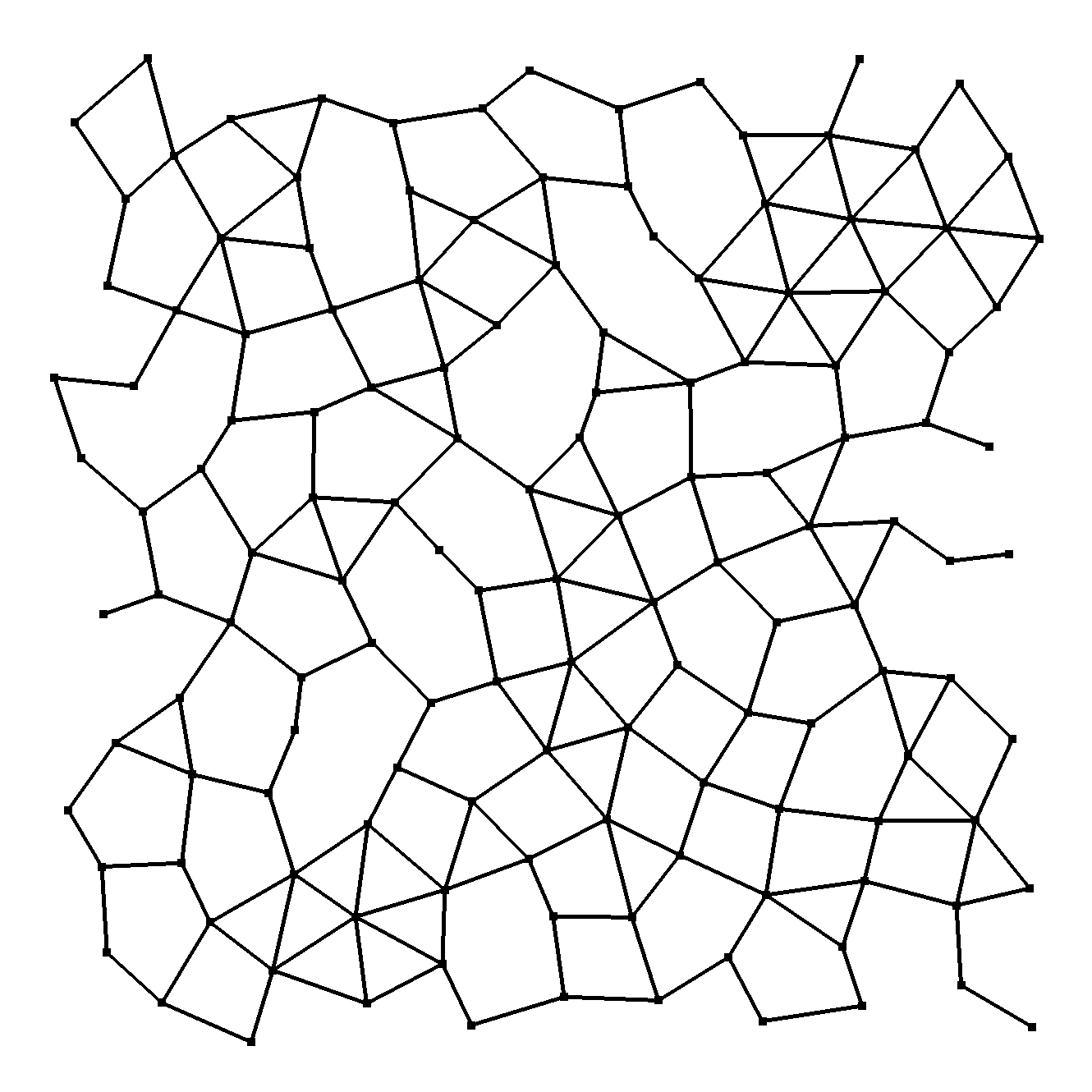}
  \includegraphics[width=0.32\columnwidth]{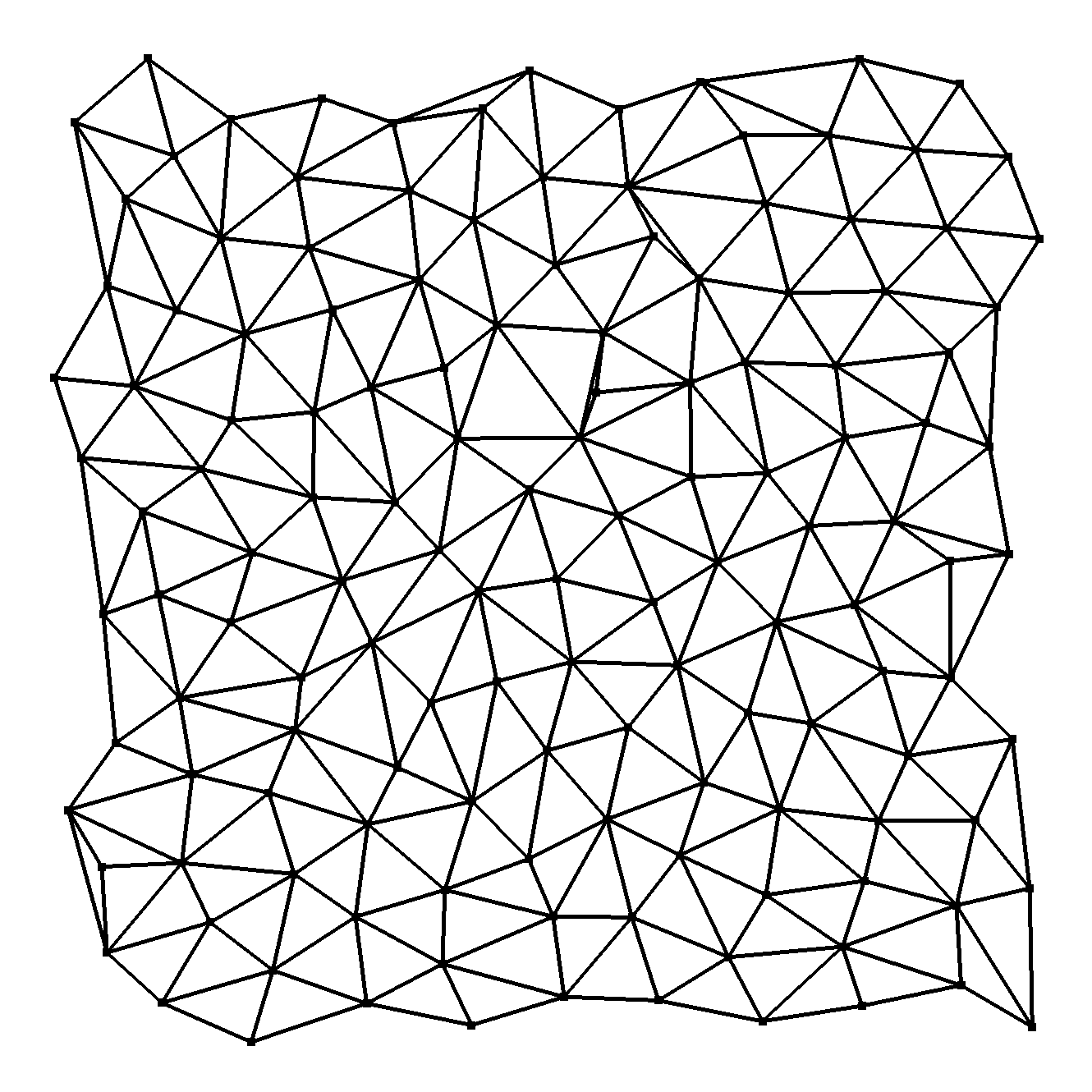}
  \caption{Reconstruction of a 2D point cloud. From left to right, the top row shows the input point cloud, the graph formed by connecting each point to its nearest neighbors within a given radius, and the minimum spanning tree of this graph. The bottom row shows the process of adding edges to the minimum spanning tree. The edges are added in order of increasing length, and edges are only added if they do not violate planarity. From left to right, the bottom row images show the MST with 10, 100, and all possible edges added.}
  \label{fig:2d_points}
\end{figure}

To obtain a triangle mesh, we recursively insert edges that split polygons into smaller polygons. The process stops when all polygons - except the boundary polygon - are triangles. A simple 2D example is shown in Figure \ref{fig:2d_points} for a synthetic point cloud.

For moderate input sizes and a 2D point cloud, the algorithm outlined above is simple to implement because the points are already embedded in the plane, and the rotation system is given implicitly. On the other hand, when designing an efficient and effective 3D version of the algorithm, we face several challenges. First of all, the rotation system, i.e. the ordering of outgoing edges for each vertex, must be computed consistently for all vertices. \revision{We refer the readers to Section~\ref{sec:init} for the definition of a consistent rotation system}. Moreover, when connecting points, several geometric considerations are important, and we want to allow connections that violate planarity in order to obtain meshes of arbitrary genus. Finally, efficiency is important: when an edge is added, we split a face, and one of the two resulting faces must be relabeled. This is a costly procedure in the initial stages of the algorithm, and the efficiency of adding edges is an important concern.

In summary, our main contributions are as follows. We
\begin{itemize}
  \item provide a rotation system-based algorithm for triangulating a 3D point cloud by iteratively adding edges to an initial spanning tree. 
  \item introduce the \textit{topology test}. This test checks whether the mesh would remain planar if a specific edge were added. The topology test is implemented efficiently, and it ensures both that spurious topological handles are not inserted and that non-manifold configurations are avoided. 
  \item allow for both topology adaptivity and control. The topology test provides us with a way to detect handle edges which we can add in order to increase the genus of the object. Since handle insertion is optional, we can also restrict the genus.
\end{itemize}

\subsection{Related Work}
%
The vast literature on methods for reconstructing surfaces from points can broadly be divided into methods which interpolate the original points and methods which approximate the point set. The former methods are often \textit{combinatorial} in the sense that they form meshes whose vertices are the input points, and the latter methods are often \textit{volumetric} (or \textit{implicit}), producing a function, $\phi: \mathbb R^3 \rightarrow \mathbb R$, which maps points in space to an intensity value or distance to surface. A good example of this type of method is the well known Poisson reconstruction method \cite{kazhdan2006poisson,kazhdan2013screened,kazhdan2020poisson} which computes a function, $\phi$, whose gradient field matches the smoothed field of estimated normals for the input points. 

Volumetric methods are generally simpler to implement than combinatorial methods, are usually extremely robust, and excel at hole closing. For a comprehensive survey of volumetric methods for surface reconstruction, the reader is referred to Berger et al. \shortcite{berger2017survey}

Our main motivation for eschewing volumetric reconstruction in favor of combinatorial reconstruction is that the former class of methods inherently smooth the reconstructed surfaces. \revision{ \sout{which leads} Admittedly, this is usually desirable, but in some cases this inherent smoothing leads} to a loss of fine detail as has been noted previously \cite{Digne2011}. By decoupling the smoothing from the reconstruction, combinatorial methods leave the smoothing to downstream processing where anisotropic methods based on e.g. fuzzy median or bilateral normal smoothing \cite{Shen04,Zheng11} or $L_0$ minimization \cite{He13}, which are designed to preserve features such as sharp edges and corners, can be brought to bear. Hence, the decoupling of smoothing and reconstruction has a tangible benefit \revision{in some cases}. This benefit is only meaningful, however, if the combinatorial methods are  robust. Hence, robustness is one of our key concerns.

\subsubsection{Combinatorial reconstruction}
The earliest combinatorial reconstruction method appears to be the work of Jean-Daniel Boissonnat \shortcite{Boissonat84}. Boissonnat proposes two algorithms, the first of which incrementally grows a polyhedron from a starting edge. The second algorithm is based on removing tetrahedra deemed exterior from the Delaunay tetrahedralization of the point cloud. The mesh resulting from this so-called sculpting process is the boundary of the union of the remaining tetrahedra. Numerous other works, notably the Crust, Power Crust, and CoCone methods of Amenta et al. \cite{amenta1998new, amenta2000simple, amenta2001power} and Edelsbrunner's Wrap algorithm \cite{Edelsbrunner2003} also build on 3D Delaunay tetrahedralization or its dual the Voronoi diagram. While methods based on 3D Delaunay tetrahedralization can provide guarantees based on sampling conditions, they are generally not robust to noise. We surmise that this is because noise in the input points invariably results in a halo of tiny tetrahedra between the interior and exterior. Extracting a reasonable surface requires selecting which triangular faces to include in the output surface. Dey and Goswami \shortcite{Dey06} provide one solution to this problem which is to select the subset of the points that lie on the boundaries of exterior Delaunay balls. However, since our aim is to include the vast majority of points, even in the presence of noise, 3D Delaunay tetrahedralization would be problematic.

Boltcheva and Levy \shortcite{boltcheva2017surface} proposed a method (Co3Ne) in which a Voronoi cell is computed for each point and restricted to a disk that is also defined at the same point. Point connectivity can then be inferred from these restricted Voronoi cells. The method requires smoothing for real-world data but it is very efficient and easy to parallelize. We compare against Co3Ne and find that it is much faster than our method but retains fewer input points and produces more holes in the output. Moreover, Co3Ne does not allow for restriction of the genus of the output shape.

More recently, Wang et al. \shortcite{Wang22} proposed a method that is likewise based on restricted Delaunay tetrahedralization but restricts to a sequence of globally defined surfaces that approach the final output. This method appears to be effective and robust to variations in sampling density, but at the cost of making a sequence of tetrahedralizations. 

We refer the reader to  Cazals and Giesen \shortcite{Cazals04} for an excellent overview of (older) reconstruction methods based on Delaunay triangulation and Voronoi cells.



Bernardini et al. \shortcite{bernardini1999ball} proposed the method known as \textit{ball-pivoting} algorithm (BPA). This method is reminiscent of Boissonnat's first method as it also grows the mesh a vertex at a time. The BPA starts from a seed triangle and uses the heuristic of pivoting a ball around an edge until it touches the next point, thereby defining a new triangle. In the presence of noise, BPA has a tendency to leave isolated points which are not touched by the rolling ball. A solution proposed by Digne et al. \shortcite{Digne2011} is to use a scale space approach where the point cloud is first smoothed and then the BPA is applied to the smoothed point cloud. This yields a reconstruction with far more of the input points, and the original vertex positions are stored and used in the final mesh. Our approach includes a smoothing step similar to that introduced by Digne et al., but it is otherwise quite different. Importantly, we include even more of the input points as demonstrated by our comparisons.

Typically, learning-based methods for surface reconstruction are volumetric in nature, but combinatorial methods exist. Notably, Sharp and Ovsjanikov \shortcite{Sharp20} proposed a method which uses neural networks to iteratively propose new triangles and decide whether to include them in the output mesh. Another recent strategy employed by Sulzer et al. \shortcite{Sulzer21} is to start from a Delaunay tetrahedralization and use a learning-based method to classify tetrahedra as being inside or outside. Focusing on the connectivity of points, Minghua et al. \shortcite{liu2020meshing} applied intrinsic/extrinsic metrics to filter out undesirable connections. More recently, Rakotosaona et al. \shortcite{rakotosaona2021learning} proposed a method called DSE meshing which learns selection of Delauany triangles from the input points. We compare against DSE meshing and find that it tends to be impossible to obtain a consistent orientation of the triangles. For a recent, quite broad, survey of deep-learning-based surface reconstruction, the reader is referred to \cite{Farshian23}.

By design, our method includes almost all input points in the reconstruction, but other researchers have created algorithms which intentionally reduce the number of points. A good example is the recently proposed method by Zhao et al. \shortcite{Tong2023}. In this work, quadric error metrics are used to cluster points such that cluster representatives lie at corners and on sharp features. The output mesh is the quotient graph with respect to this clustering.


The approach that is most similar to ours is the work by Robert Mencl \shortcite{Mencl95} which was later extended by Mencl and M{\"u}ller \shortcite{Mencl98}. Their starting point is also a minimum spanning tree, but in their case the \textit{Euclidean MST} is employed whereas we use the MST of a k-nearest neighbor graph. Mencl and M{\"u}ller also connect vertices and ultimately triangulate the graph, but they do not store the faces of the graph and, therefore, \rev{could} not implement a test similar to our topology test. This means that spurious topological handles could be created since it is impossible to tell whether an inserted edge splits a face or connects different faces. For relatively noise-free point clouds this may not be much of an issue, and, unfortunately, noisy point clouds are not considered. In later work, the same authors propose an improved method based on the so-called $\beta$ environment graphs rather than minimum spanning tree \cite{Mencl03}. While the improved method is clearly effective for clean point clouds, we are again concerned that topological errors could result from noise. 

\subsubsection{Normal estimation}
Like many surface reconstruction methods, our technique requires that consistently oriented normals are associated with the input points.
Fortunately, this is in itself an extensively researched area. Most of the early normal estimation methods apply plane-fitting techniques, e.g. \rev{Principal} Component Analysis (PCA) \cite{hoppe1992surface, mitra2003estimating, pauly2002efficient, huang2009consolidation}, more complex surface fitting methods such as Moving Least Squares (MLS) \cite{alexa2001point, levin1998approximation}, or jet fitting \cite{cazals2005estimating}. Recently, learning-based methods have been applied to normal estimation by directly regressing normal vectors \cite{guerrero2018pcpnet} or fitting surfaces \cite{ben2020deepfit, zhu2021adafit, li2022hsurf}. For our purposes, the older methods (e.g. \cite{hoppe1992surface} are sufficient, and in most cases we just use the normals supplied with the test data. However, we also include a test to elucidate the effect of normal noise. 

\subsubsection{Rotation systems}
John Robert Edmonds \shortcite{Edmonds1960} showed that for any graph imbued with a cyclic ordering of the incident edges for each vertex, i.e. a rotation system, there is a corresponding, topologically unique polyhedron\footnote{Edmonds work was independent of, but much later than, the work of Lothar Heffter \shortcite{Heffter1891} on which Gerhard Ringel  \shortcite{Ringel1965} built.}. 
In the context of interactive modeling, Akleman and Chen \shortcite{Akleman99} used rotation systems to design the principles of their {doubly linked face list} (DLFL) representation for polyhedral shapes.
Later, Akleman et al. \shortcite{Akleman2003} presented a DLFL-based system where meshes are built using a minimal set of operators which includes an operator that creates a handle by adding an edge connecting two distinct faces.

\section{Definitions}
\revision{In this section, we provide a number of formal definitions which are necessary for the exposition of the method.}

\revv{An initial step of our approach is to form a \textit{spatially embedded graph} from the 3D input points.} We define a graph, $G$, as the tuple $G=(V, E)$, where $V$ is a set of \textit{vertices} and $E$ is a set of \textit{edges}. \revv{In the following, we will assume that a point, $\mathbf p_v \in \mathbb R^3$, is associated with each vertex $v \in V$.} An edge, $e = \{v,w\}$ where $v,w \in V$, is an unordered pair of vertices. For our purposes, loops where $v$ and $w$ are the same vertex will not be needed, hence $v \ne w$.

\revision{We define a \textit{genus-0} graph as a graph that \textit{can be} embedded in a manifold of genus 0 or higher such that edges intersect only at vertices. A genus-0 graph is sometimes called a planar graph since it can be drawn in the plane, but we use the term genus-0 to emphasize that an embedding surface that interpolates the vertices may be far from planar. Our method relies on the fact that if a graph is a tree, it is necessarily a genus-0 graph. This follows from the observation that starting from a single node, we can draw any tree in 2D one branch at a time such that every new branch is drawn without crossing a previously drawn branch.}


\revision{A \textit{plane projectable} graph (or pp graph) is a graph for which there is a \revv{3D unit vector, $\mathbf d \in S^2$}, such that when \revv{the points associated with the vertices of the graph are} projected into 2D along \revv{$\mathbf{d}$}, no two edges intersect except at \revv{the projected points}. In this paper, we will only require that small sub-graphs of a given graph are pp graphs.}

\subsection{Rotation Systems}
\label{sec:face_def}
\revision{In order to define rotation systems, we first need the set of \textit{halfedges}}, $H = \{ (v,w)\; | \; v,w \in V \wedge \{v,w\} \in E \}$. If there is an edge between vertices $v$ and $w$, then $(v,w)$ and $(w,v)$ both belong to $H$. Thus, halfedges, which are sometimes referred to as \textit{darts}, are simply the directed versions of edges. If $v$ is the first vertex in a halfedge $(v,w)$, we say that $(v,w)$ is \textit{outgoing} from $v$.

\begin{figure}[ht]
    \centering
    \includegraphics[width=0.7\columnwidth]{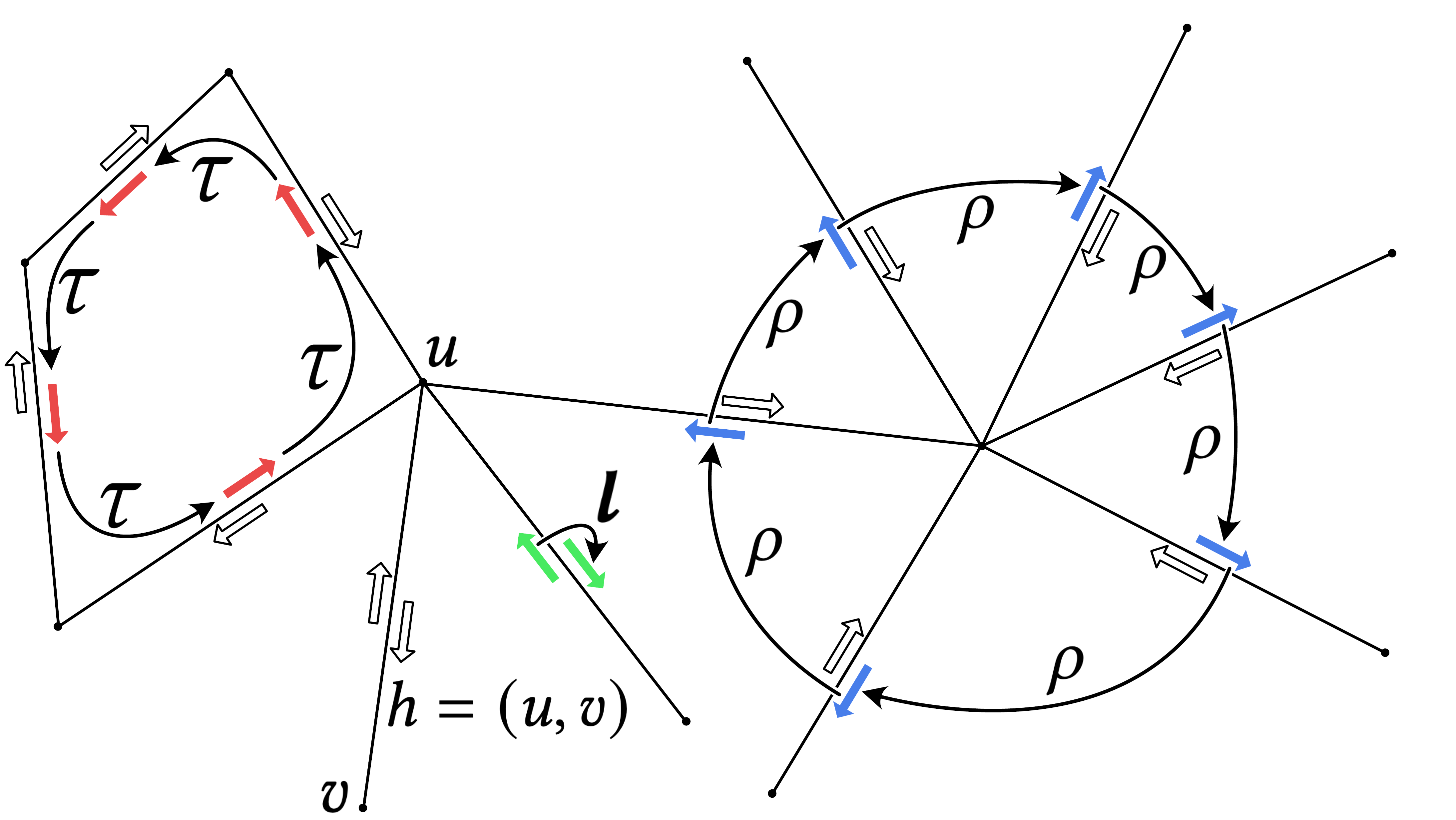}
    \caption{This figure shows a partial graph consisting of vertices, edges, and halfedges (ordered pairs of vertices) indicated as fat arrows. The functions $\rho$, $\iota$, and $\tau = \rho \circ \iota$ map halfedges to halfedges. The thin arrows show the actions of the functions. Orbits of the functions are shown in red (faces), \rev{green} (edges), and \rev{blue} (one ring).}
    \label{fig:operators}
\end{figure}

We will define a polygonal mesh\revision{\sout{, $M=(V, E, F)$,}} as a surface constructed from a graph by associating a set of faces, $F$, with the graph. As described by Edmonds~\cite{Edmonds1960}, we can obtain the faces of a closed, two-sided polygonal mesh by associating a \textit{rotation system} with the graph. A rotation system assigns cyclic orderings of the halfedges outgoing from $v$ for all $v\in V$. We can define the rotation system in terms of a bijective function $\rho : H \rightarrow H$ which maps an outgoing halfedge, $h=(v,w)$ to the next outgoing halfedge in the cyclic ordering around $v$ as shown in Figure~\ref{fig:operators}. 
In some cases, a vertex, $v$ has only a single incident edge, and then $\rho((v,w))=(v,w)$ where $w$ is the only neighbor of $v$.

In order to define the faces of a mesh in terms of a rotation system, we need the bijection $\iota: H \rightarrow H$ which inverts the orientation of the halfedge. In other words, if $h=(v,w)$ then $\iota(h)=(w,v)$. This means that $\iota$ is a \textit{fixed-point free involution}, i.e. $\iota(\iota(h))=h$ and there is no halfedge $h$ left invariant by $\iota$.

A \textit{face is} defined in terms of the composition $\tau = \rho\circ\iota$. \revision{Specifically, $\tau$ maps a halfedge to the next halfedge in the cycle that bounds a face, and} the face associated with a halfedge, $h$, is $f_h = \{\tau^n(h)| n \in \N_0\}$ or, in other words, the \textit{orbit} of $h$ under $\tau$. The set of all faces can now be defined as $F = H/\sim$ where $\sim$ is the equivalence relation defined by $h \sim h'$ if and only if $h' \in f_h$. \revision{We note that in our method we define faces directly in terms of the rotation system. Only when the triangles are ouput do we convert the mesh to an indexed face set.}

A \textit{corner} of a face, $f$, at a vertex, $v$, is given by a pair of outgoing halfedges $h$ and $h'$ which are adjacent in the cyclic ordering, i.e. $\rho(h)=h'$ and such that $h'\revision{, \iota(h)} \in f$. A corner can be split by inserting a new outgoing halfedge between the two adjacent halfedges in the cyclic ordering.

In the same vein as faces can be defined as orbits of $\tau$, we can see edges as the orbits of $\iota$ and vertices as the orbits of $\rho$. The actions of $\rho$, $\iota$, and $\tau$ along with their orbits are illustrated in Figure~\ref{fig:operators}.
\subsection{Euler Operators}
\label{sec:euler_op}
\begin{figure}[h]
    \includegraphics[width=\columnwidth]{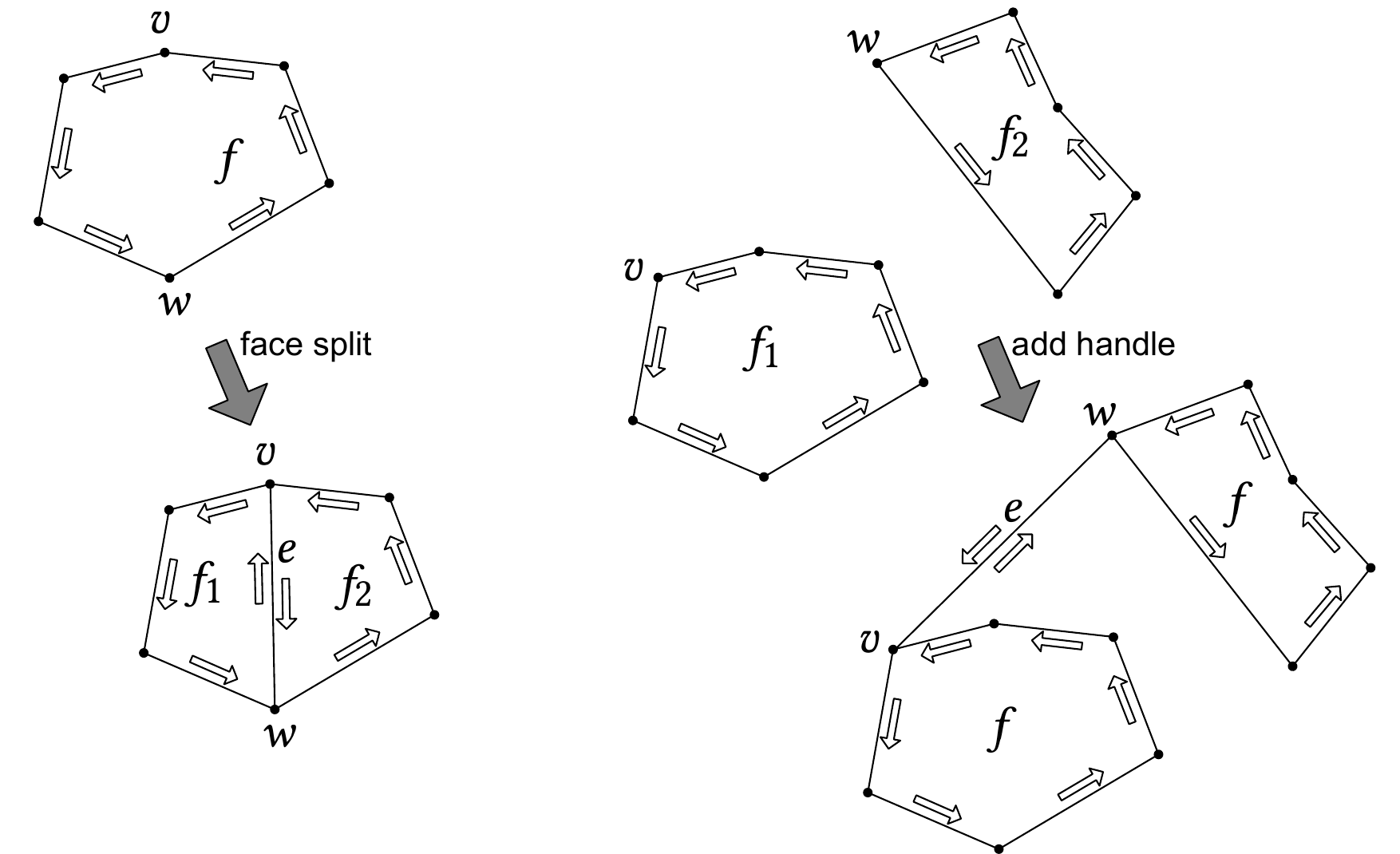}
    \caption{Before (top) and after (below) illustration of the edge insertion operation both in the case (left) where a face $f$ is split into two faces and in the case (right) where two faces, $f_1$ and $f_2$, are merged into a single face $f$. Note that the new face, $f$, is a tube that connects the boundaries of $f_1$ and $f_2$ which is hard to convey adequately in a 2D drawing.}
    \label{fig:edge_insertion}
\end{figure}

Editing operations on a manifold mesh, $M=(V,E,F)$, are often defined in terms of the Euler-Poincare formula,
\begin{equation}
    |V|-|E|+|F| = 2(1-g) \enspace ,
    \label{eq:euler}
\end{equation}
where $g$ is the genus of the object represented by the mesh, and the value $2(1-g)$ is known as the Euler characteristic. This formula must be satisfied by any closed, manifold mesh. As such, it has been used as an invariant in solid modeling operations as discussed by e.g. Mantyla and Sulonen \shortcite{Mantyla1982}. Typically, operators that maintain the Euler-Poincare formula invariant are called \textit{Euler operators}. In the following, we will describe the two Euler operators we will need for our surface reconstruction algorithm and how the rotation system of the mesh is affected by these operations. The operators are illustrated in Figure~\ref{fig:edge_insertion}.

\subsubsection{Splitting a face by edge insertion}
\label{sec:def_edge_insertion}
A face is split by creating an edge, $e$, between two non-adjacent vertices, $v$ and $w$. Assuming $v$ and $w$ belong to the same face, $f$, this operation splits $f$ into two faces. Thus, it is clear that \eqref{eq:euler} holds since $|E|$ and $|F|$ both increase by one. 

If we define our mesh in terms of a rotation system, the cyclic orderings of outgoing halfedges need to be updated for $v$ and $w$. Moreover, the corners of $v$ and $w$ which are split by the insertion of $(v,w)$ and $(w,v)$, respectively, must both be associated with $f$ for the edge insertion to be valid.

\subsubsection{Adding a handle by edge insertion}
A topological handle is added by creating an edge, $e$, between a vertex $v$ that belongs to face $f_1$ and a vertex $w$ belonging to a different face, $f_2$. In this case, $f_1$ and $f_2$ are merged into a single face which we can see as a tube connecting the holes left by removing $f_1$ and $f_2$. Intuitively, one can think of the new face as a piece of paper that has been rolled into a cylinder: the boundary curves of the cylinder correspond to the boundaries of $f_1$ and $f_2$ whereas $e$ corresponds to the edge where the opposite sides of the paper meet.

Topologically, this operation adds a handle to the mesh, increasing $g$ by one. Equation \eqref{eq:euler} still holds since $|E|$ and $|F|$ increase and decrease by one, respectively.

The same caveat applies for handle insertion as for face splitting. At vertex $v$ we need to add the $(v,w)$ halfedge to the cyclic ordering of outgoing halfedges in a corner that belongs to $f_1$ and at vertex $w$ we need to add the $(w,v)$ halfedge in a corner that belongs to $f_2$.

\begin{figure}[]
    \centering
    \includegraphics[width=0.4\columnwidth]{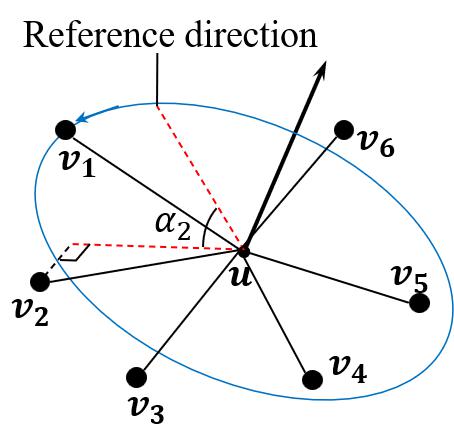}
    \caption{$RS$ of an example neighborhood. Taking $v_2$ as an example, it is projected on the disk. Then an angle $\alpha_2$ is calculated as the radian of vertex $v_2$.}
    \label{fig:RS}
\end{figure}
\section{Method}
\label{sec:method}

\begin{figure*}[]
    \centering
    \includegraphics[width=\textwidth]{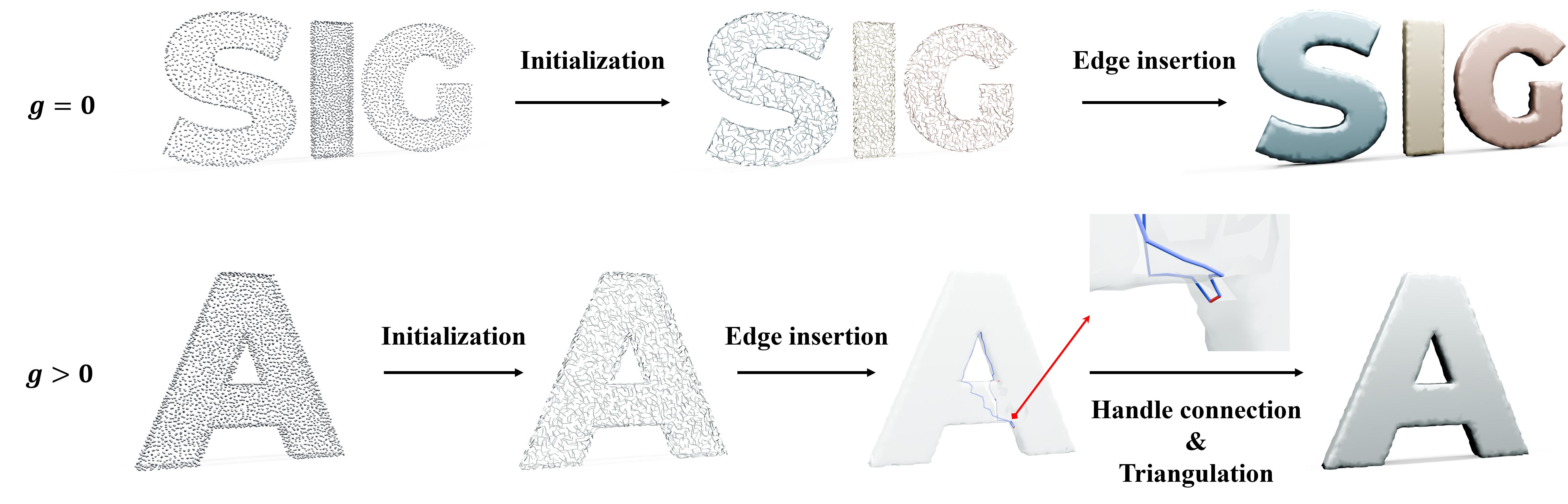}
    \caption{Pipeline overview on synthetic letter point clouds. \revision{\sout{General} High-genus} reconstruction follows the bottom pipeline, where $g$ stands for the genus number. \rev{Given the prior knowledge of a genus-0 shape, the shorter pipeline depicted at the top is employed. In the pipeline shown at the bottom, the red edges in the third inset denote the newly added handles. Concurrently, the path formed by the blue edges represents the shortest path between the two end vertices before the red edge is introduced.}}
    \label{fig:pipeline}
\end{figure*}

The essence of our algorithm is to construct a tree from the input points and build a triangle mesh from this starting point. A\revision{s mentioned, a} tree is known to be \revision{\sout{a planar graph} genus-0}, and we ensure that \revision{\sout{planarity}this property} is maintained while \rev{regular} edges are added. As an optional part of the algorithm, handle edges can be added to increase the genus of the reconstructed object. 

In the following, we will discuss all stages in greater detail and we refer the reader to Figure~\ref{fig:pipeline} and Algorithm~\ref{alg:main} for an overview of our pipeline.

\begin{algorithm}
    \caption{\revision{Pseudocode for the main loop}}
    \label{alg:main}
    \SetKwInOut{Input}{Input}
        \Input{pts, $r, k, \theta, n$, normals}
        pts\_proj $\gets$ pts\\
        \If{points are noisy} {
            \rev{pts\_proj $\gets$ project(pts, normals)} // c.f. Section~\ref{sec:projection_dist}\\
        }
        \rev{$\Gamma \gets$ build\_knn\_graph(pts\_proj, $r$, $k$)}\\
        \For{$G$ \textbf{in} components($\Gamma$)}{
            $CO \gets$ build\_circular\_ordering($G$, normals)\\
            MST $\gets$ init\_MST($G$)\\
            $(M, RS) \gets$ (MST, build\_RS(MST, $CO$))\\
            $Q \gets$ sort\_edges($G$)\\
            \For{edge \textbf{in} $Q$}{
                \If{$\mathrm{topology\_test}(M, RS, \mathrm{edge})$}{
                    \If{$\mathrm{geometry\_test}(M, RS, \mathrm{edge}, \theta,$ \rev{pts\_proj})}{
                        \rev{$\mathrm{insert\_edge}(M, \mathrm{edge})$}\\
                        $RS \gets \mathrm{update\_RS}(RS, CO$, edge)\\
                    }
                }
            }
            \If{handle connection enabled}{
                $\mathrm{connect\_handle}(M, RS, n)$\\
                $\mathrm{triangulate}(M, RS)$\\
                \rev{// c.f. Section~\ref{sec:handle}}
            }
            output\_mesh($M, RS$, pts)\\
        }
\end{algorithm}

\subsection{Initialization}
\label{sec:init}
\revision{Given the input point cloud, we build a k-nearest neighbor ($k$NN) graph, $\Gamma$, where $k$ is a user-defined parameter and the Euclidean distance is used to define the neighborhood. Notably, since the connections are symmetrized for both incident vertices of an edge, a vertex may have more than $k$ neighbors. However, edges are also removed if the connected vertices $u$ and $v$ have normal vectors which point in significantly different directions, i.e. $\mathbf N_u \cdot \mathbf N_v < cos(\theta)$, where $\theta$ is also a user-defined parameter. This test helps us avoid connecting vertices on either side of thin structures. Finally, edges longer than $r$ times the average edge length are removed as outliers. We refer the reader to Section~\ref{sec:parameter} for further discussion of $r$ and the two other parameters.}

\revision{For input point clouds that are known to be free of noise, we can use $\Gamma$ directly in the reconstruction process described below. However, for noisy inputs, we compute a filtered point set and a corresponding graph as described in Section~\ref{sec:projection_dist}.}

\revision{The next step is to compute and iterate through all the connected components of $\Gamma$ and reconstruct a triangle mesh for each component, $G=(V,E_G)$.}


\begin{figure}[]
    \centering
    \includegraphics[width=0.6\columnwidth]{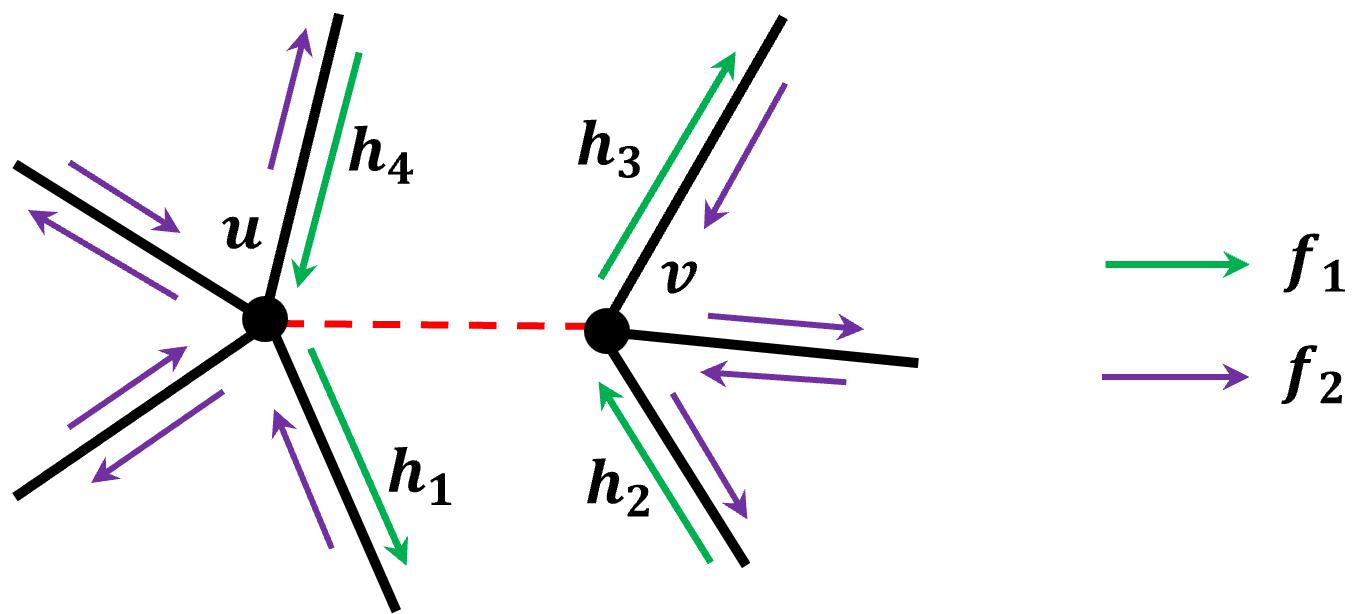}
    \caption{Topology check for the candidate edge $\left\{u,v\right\}$. \revision{In this example, the candidate edge splits $f_1$ within the right corner. Thus, it passes the topology check.}}
    \label{fig:topocheck}
\end{figure}

\revision{Based on the vertex normals\sout{At this time}}, we now establish a cyclic ordering\revision{, $CO$,} of the edges incident on all vertices of $G$. \revision{\sout{The construction of the cyclic ordering around an arbitrary vertex, $u$,} The construction of this ordering for a vertex $u$} with neighbors, $v_i$, \revision{where $\left\{u,v_i \right\} \in \sout{E} E_G$} is shown in Figure~\ref{fig:RS}. Each neighboring vertex, $v_i$, is projected into the plane defined by the position $\mathbf{p}_u$ and the normal, $\mathbf N_u$, of vertex $u$. We compute the angle of each projected vertex relative to an arbitrary reference direction and thus obtain a cyclic ordering of the neighbors of $u$.
\revision{\sout{which is finally stored in $RS$.}}
While our approach thus relies on vertex normals, the normal computation is not an intrinsic part of the scheme, and our code allows for user-provided normals. In both respects, our approach is similar to many other reconstruction methods. 

As the next step, a graph, \revision{$M=(V, E_M)$, is initialized to the minimum spanning tree (MST) of $G$, and we initialize the rotation system, $RS$, such that the edges $E_M \subset E_G$ have the same cyclic order in $RS$ as they do in $CO$. Only if this is true do we consider the rotation system to be \textit{consistent} with the input normals.}

\revision{At this point, $M$ defines the vertices and edges of a mesh, and $RS$ defines the faces in terms of the operator $\rho$. Since $M$ is a tree initially, it consists of only a single face} that can be traversed starting from an arbitrary halfedge using repeated application of $\tau = \rho \circ \iota$. \revision{In the next phase, additional edges from $E_G$ will be added to $E_M$. As $E_M$ expands, $RS$ will be updated while ensuring that it remains consistent as defined above.} 
\revision{\sout{Of course, $M$ does not contain all edges of $G$, and at any time the cyclic ordering is restricted to the subset of halfedges included in $M$.}}

In addition to the rotation system, we need to be able to keep track of the face associated with each halfedge. A balanced binary tree (not to be confused with the minimum spanning tree of $G$) is used for this purpose. From the tree, the face associated with each halfedge can be retrieved, and we can update the tree efficiently when a new edge is inserted, splitting one of the faces into two faces (for more information, please refer to Section~\ref{sec:Face}). 


\revision{
As the final initialization step, all edges in $G$ are sorted in order of increasing length and stored in the list $Q$} \rev{since we will consider edges in this order}.


%
\subsection{Edge Insertion}
\label{sec:edge_insertion}
In the edge insertion stage, we iteratively \revision{pick edges from $Q$ and check whether they are valid}. \revision{Valid edges\sout{$G$ $E_G$ into the mesh $M$} are inserted into $E_M$}, thereby splitting a face of $M$ as illustrated in Figure~\ref{fig:edge_insertion}.
\revision{\sout{The edges are sorted according to length, and an} An} edge is considered valid \revision{\sout{for $M$}} if it passes two tests, the topology test and the geometry test, which, \revision{respectively}, ensure that $M$ \revision{\sout{remains a valid} remains a genus-0 graph and locally a pp graph. In other words, the first test is required to ensure that we do not make any topological errors and the second test ensures we avoid flipped triangles and overlapping edges in the mesh.}


\revision{\sout{Recall that the edges of a genus 0 mesh constitute a {planar} graph.  A graph is planar if it can be embedded in 2D in such a way that no two edges intersect except at vertices. In the following, when we refer to $M$ as being planar, it means that the edges of $M$ form a planar graph. It is also important to distinguish between planar and plane graphs. A plane graph is a graph that is actually embedded in 2D without any edges crossing each other. Hence, a plane graph is necessarily a planar graph, but the converse is not true. }}

\revision{\sout{The first of our two tests, which we \revision{\sout{denote} refer to as} the topology test, ensures that the mesh remains \revision{\sout{planar} genus-0} if the edge candidate is inserted. \revision{\sout{This may seem sufficient, but in some cases, a planar graph can be difficult to embed in 2D despite (in fact) being planar.} However, it doesn't get rid of potential intersections of triangles. For this reason, we also carry out the heuristic geometry test which checks that \sout{edges do not cross locally}it suffices a pp graph within a defined neighborhood of any vertex.} Both of these tests are described in the following.}}
\subsubsection{Topology Test}
\label{sec:topotest}
\revision{\sout{The topology test ensures that the entire graph remains planar when a new edge is added.}} This test first ensures that the endpoints of the candidate edge belong to the same face\revision{\sout{.},} $f$. Next, it is verified that both of the corners which are split when the candidate edge is inserted belong to $f$. \revision{We refer the reader to Section~\ref{sec:face_def} for the definition of corner. \sout{If this test is not passed, we reject the candidate edge.}}

In order to carry out the topology test in practice, we store the face associated with each halfedge. When a candidate edge $\left\{u,v\right\}$ is considered, we check that all the halfedges before and after $\{u,v\}$ in the rotation system, i.e. $h_1,h_2,h_3$ and $h_4$, are associated with the same face, $f_1$ \revision{as shown in Figure~\ref{fig:topocheck}.}

\begin{figure}[]
    \centering
    \includegraphics[width=1.0\columnwidth]{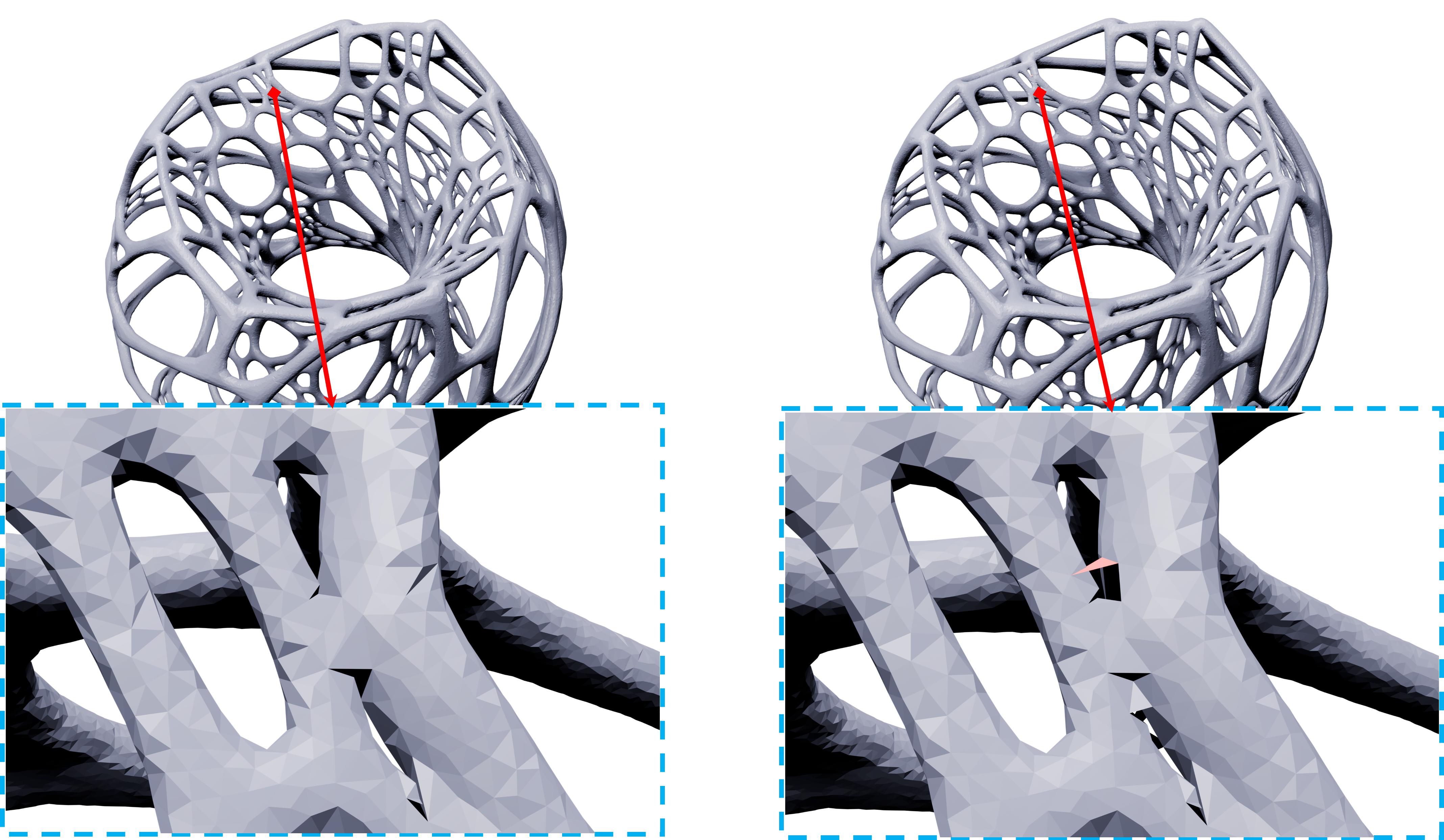}
    \caption{Example for the necessity of topology test. The left mesh is reconstructed with both tests, while the right one is reconstructed without \revision{the} topology test. A red triangle is formed which result\revision{s} in an undesirable hole.}
    \label{fig:ab1}
\end{figure}

It is clear that $M$ remains \revision{\sout{planar} genus-0} if the edge candidate passes the topology test because $|E|$ and $|F|$ both increase by one. Consequently, the left\revision{-}hand side of the Euler-Poincare formula \eqref{eq:euler} does not change. This also implies that \revision{\sout{the}} the topology test ensures that handles are not inserted at this stage, \revision{\sout{and this is the reason} which is why our method enables reconstruction} with topology control. 
\subsubsection{Geometry Test}
\begin{figure}[h]
    \centering
    \includegraphics[width=0.6\columnwidth]{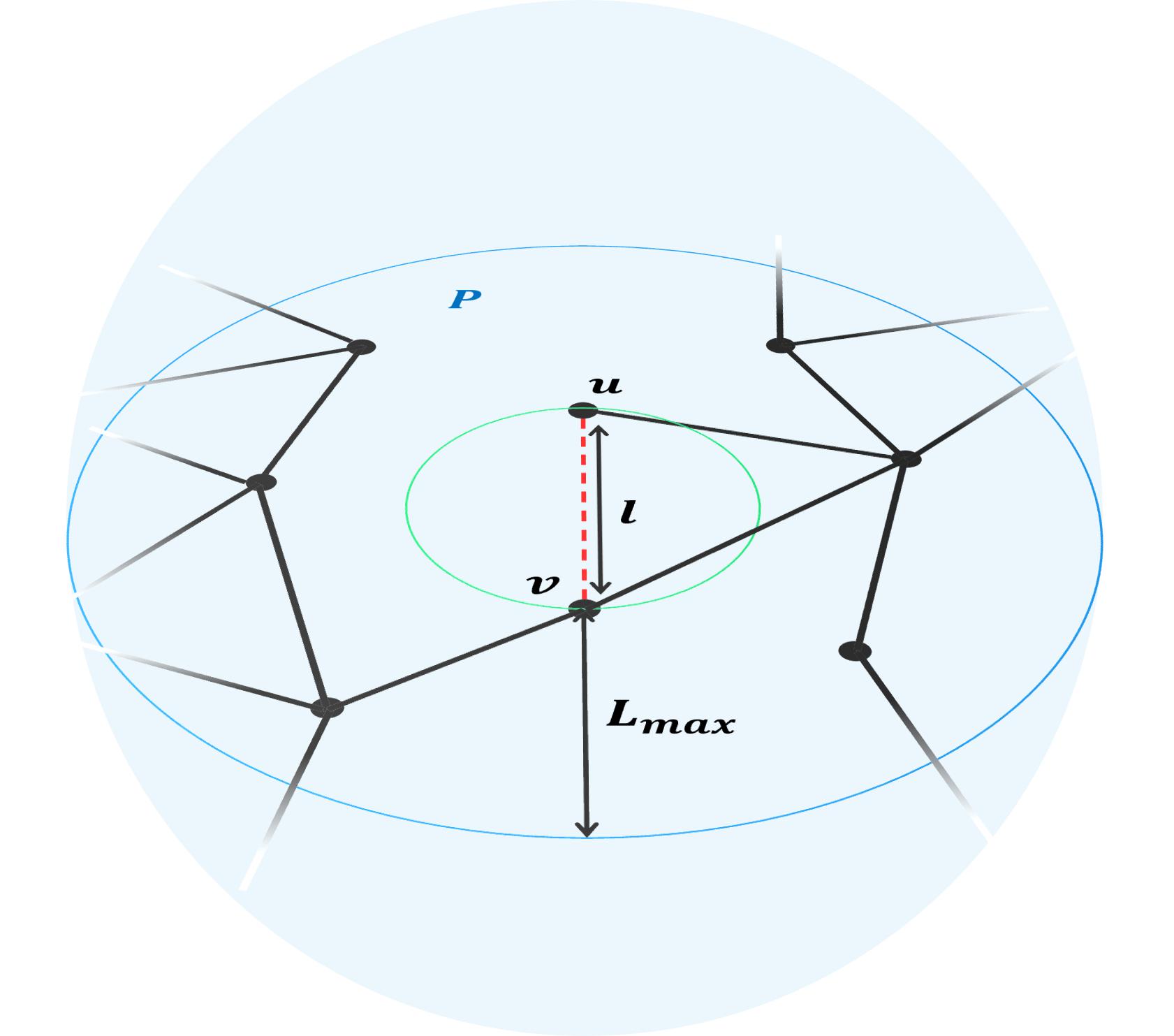}
    \caption{Geometry test for edge \{$u,v$\}. \revision{With a search ball of radius $\frac{1}{2} l + L_{\max}$ and centered at the midpoint of the edge candidate, this test will check if any existing edge associated with vertices inside this search ball intersects with the candidate edge within the tangent plane.}
    }
    \label{fig:geom_test}
\end{figure}

The geometry test ensures that a 2D projection of a local neighborhood of the graph remains a plane graph, \revision{i.e. a pp graph}, when a new edge is added. As depicted in Figure~\ref{fig:geom_test}, the local neighborhood is defined as a ball with a radius of $\frac{1}{2} l + L_{\max}$, where $l$ is the length of the edge candidate and $L_{\max}$ is the maximum length of any edge in $G$. The plane, $P$, passes through the midpoint of the candidate edge and its normal is defined as the average of the normals at the endpoint vertices.

Any edge for which it is true that either endpoint belongs to the local neighborhood is projected into $P$ and we test for intersection with the candidate edge. If there is an intersection, we reject the candidate edge.

\revision{\sout{We can understand the geometry test as ensuring that $M$ remains a plane graph locally. However,}}

\revision{We stress that both the topology test and the geometry test are required. \sout{ In regions where the normal changes rapidly, the geometry test can be insufficient to guarantee that $M$ remains a planar graph as shown in Figure~\ref{fig:ab1}. Hence, we need the topology test. it is important to stress that}} Even if topology control is not desired, the topology test is needed in cases where the tangent plane changes rapidly. In such cases, the geometry test may pass for edges that connect different faces as illustrated in Figure~\ref{fig:ab1}.

\begin{figure}[h]
    \centering
    \includegraphics[width=1.0\columnwidth]{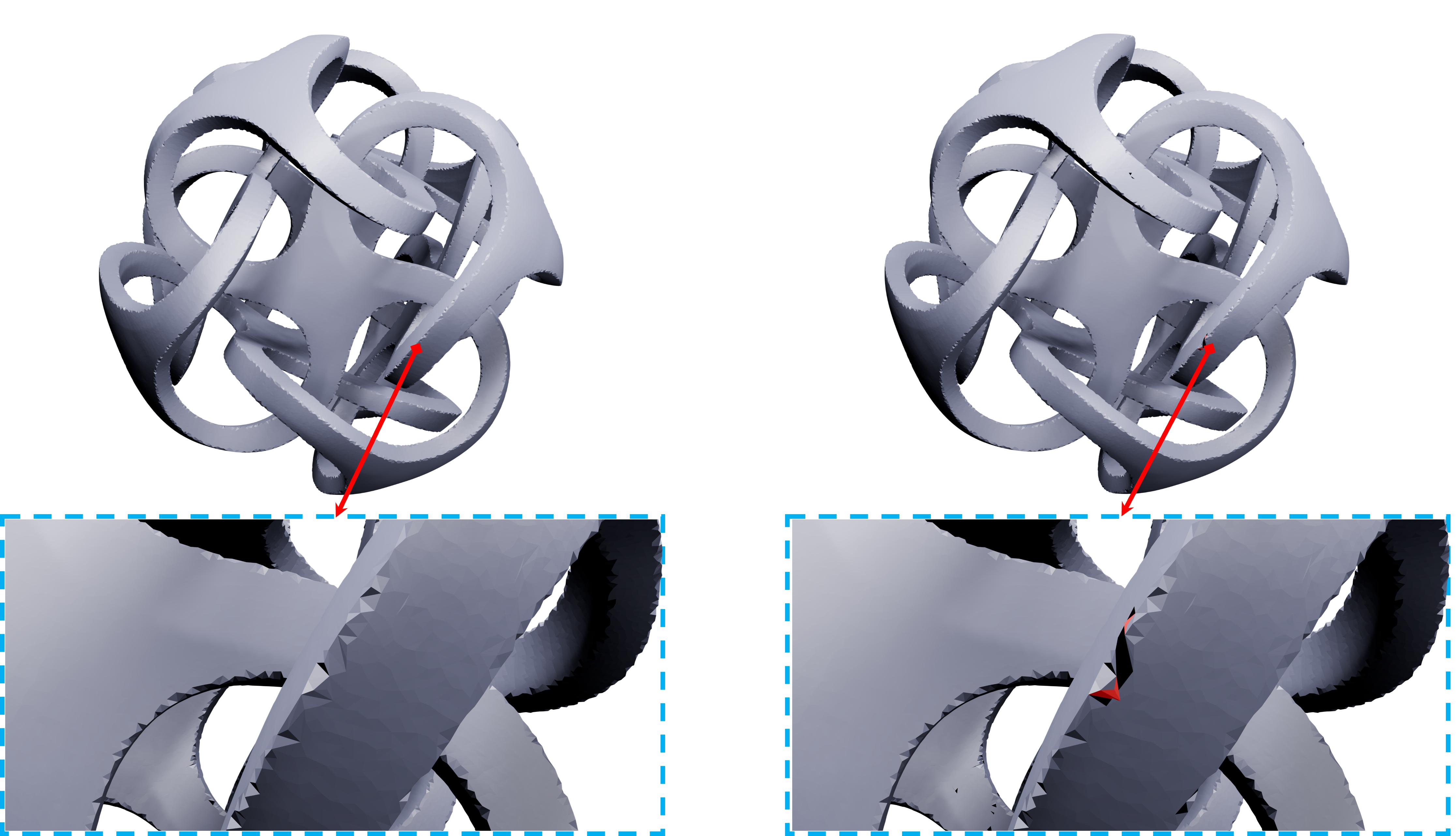}
    \caption{Example for the necessity of geometry test. The left mesh is reconstructed with both tests, while the right one is reconstructed without \revision{the} geometry test. The \revision{graph remains genus-0\sout{local graph of the mesh is still planar}}, but \rev{no} edge candidate left can pass the corner check of \revision{the} topology test. Thus, a hole is left here and the same reason for any other place of the reconstructed mesh.}
    \label{fig:ab2}
\end{figure}

\revision{\sout{On the other hand, just like the geometry test} Likewise}, the topology test is not sufficient in itself. As shown in Figure~\ref{fig:ab2}, there are cases where an edge that passes the topology test would lead to \revision{undesirable configurations, such as intersections, \sout{that are highly undesirable}}\revision{which do not increase the genus number\sout{cause the graph to be non-planar}. \sout{The reason \revision{\sout{for this}}is that a graph can be planar yet very far from being plane.}} Fortunately, these cases are easily dealt with by the geometry test. 

Candidate edges \revision{\sout{which} that} pass both the geometry and topology test can safely be inserted. However, the two tests do not guarantee the quality of the triangles. In order to prevent skinny triangles, i.e. caps or needles with one angle close to $180^\circ$ or $0^\circ$, respectively, we also perform a quality check before insertion. This is discussed in Section~\ref{sec:quality_concern}.

\revision{It bears mentioning that both the topology test and the geometry test depend on the normal vectors. In the case of the topology test this means that very noisy normal vectors could result in fewer edges that pass the test, but it does not result in edges erroneously passing the test. In the case of the geometry test, no strong guarantees are provided, but normal vector noise would generally also lead to edges failing this test. Ultimately, this means that the deleterious effects of noisy normal vectors can be summed up as being fewer connections which, for significant levels of noise, can lead to holes in the mesh. We investigate the effect of synthetic noise on the normal vectors in Section~\ref{sec:normal-noise-test}. 
}


\subsection{Handle Connection \& Triangulation}
\label{sec:handle}

Up to this point, $M$ is guaranteed to be of \revision{\sout{genus-0, i.e. of}}sphere topology, but some of the faces of $M$ likely consist of many edges. Given such a large face, $f_1$, and a vertex, $v$, that belongs to $f_1$. $v$ may either lack connections to other vertices of $f_1$ because there is a gap in the input point cloud or because the unused edges of $G$ which are incident on $v$ connect it to a different face, $f_2$. In the former case, we would want to leave $f_1$ as a hole, but in the latter case, we may want to insert a handle edge between $f_1$ and $f_2$


To merge $f_1$ and $f_2$, we need to close the \textit{cracks}. A crack is constituted by a pair of faces $f_1$ and $f_2$ such that halfedges on one side are associated with $f_1$, while halfedges on the other side are associated with $f_2$. We create a set of handle edges by finding vertices belonging to a face $f_1$ that are connected to a different face and whose corner (i.e. the corner that would be split if the edge were inserted) has an angle greater than $\pi$. \rev{This restriction on corner angle aims to avoid skinny triangles as a quality concern. A summary of quality concerns is provided in Section~\ref{sec:quality_concern}} \sout{For further explanation of the choice of $\pi$, please refer to Section~\ref{sec:quality_concern}.} Such handle edges are now inserted if they pass the geometry test described above. The topology test does not apply since it would trivially fail for handles. On the other hand, small, undesired handle candidates are often present locally -- in particular on noisy meshes. \revision{\sout{For this reason} Thus}, we need a different test for handle insertion\revision{, and the following heuristic test is employed}.

\revision{\sout{In light of the fact that handles represent global structures, Through extensive experiments, we found handles often represent global structures, which means
the vertices incident on them usually have a short Euclidean distance but are connected by a long shortest path within the graph.}} 
\revision{When a handle edge is considered for insertion, we can find the shortest path between the endpoints that consists of edges already in $M$. If the handle is well-sampled, this path consists of a significant number of edges. For instance, consider the letter A in the lower row of Figure~\ref{fig:pipeline}: it is clear that the path connecting the endpoints of the handle edge seen in the inset must go around the hole in the A.} Thus, we use a threshold, $n$, on the number of edges in the shortest path between the incident vertices of the handle candidate edge, to cull undesirable handles. A higher value of $n$ corresponds to fewer handles being connected. \revision{\sout{In our implementation}For the input point clouds} we tested, an effective setting for $n$ is $50$. 

After connecting all handle edges that pass the above test, a final triangulation is performed. At this point, most of the faces are indeed triangles, but since we reject some edges due to quality concerns both during edge insertion and handle insertion (c.f. Section~\ref{sec:quality_concern}), not all faces are fully triangulated. After handle insertion, our face data structure is no longer valid since it can only be efficiently updated when faces are split and not merged. Fortunately, the rotation system is unaffected, and the triangulation only performs ear clipping, i.e. it connects two vertices known to belong to the same face. More details are provided in Section~\ref{sec:add_face}.
\subsection{Reconstruction from Noisy Point Clouds}
\label{sec:projection_dist}
Scanned point clouds are inherently noisy since scanning is a \rev{physical} measurement process. Moreover, several sub-scans are often combined during reconstruction. These may not be perfectly aligned, introducing another source of high-frequency noise.


Noise can be both tangential and normal to the surface. While tangential noise does not affect the reconstruction, normal noise can result in large Euclidean distances between points that are close in the tangent plane. \revision{To mitigate this issue, two measures are taken against \sout{specified}noisy data.}

\revision{First, we use the provided normals to project all points onto their tangent planes and compute the projected distances which are then used instead of Euclidean distance. To define the projection distance, we need a preliminary definition: for an edge $e = \{u,v\}$, $e_{//v}$ refers to the projection of $e$ on the tangent plane of $v$. The same applies \sout{for} to} $u$. The projection distance $|e|_p$ is now,
%
\begin{equation}
    \label{eq:projection_dist}
    |e|_p = (|e_{//v}| + |e_{//u}|)/2 \enspace .
\end{equation}
However, the kd-tree used for computing the initial graph $G$ is, of course, unaffected by the tangential distances and still based on the Euclidean distance. This can lead to both spurious and missing connections. Therefore, we also project the input points to the average tangent plane of their k-neighborhood and rebuild the kd-tree. Inspired by \cite{Digne2011}, we restore the input points to their original position before the mesh is output. 

It is reasonable to ask why we do not simply compute the Euclidean distance after projecting the points. The reason is that \revision{
even after the above smoothing, two connected points never lie exactly in each others tangent plane. Hence, there is a meaningful difference between the Euclidean distances and the tangent plane projected distance}\revision{\sout{the projected points still retain variation along the normal direction}}, and we obtain the best results by using the pure tangent plane distance.

\section{Implementation}
\label{sec:implementation}
In this section, we discuss the practical aspects of the method\sout{in the order of the pipeline outlined in Section \ref{sec:method}}. These aspects include some details \revision{\sout{at the initialization stage,} pertaining to} the binary tree data structure used to store faces, some quality concerns, details related to triangulation after handle connection, the time complexity of each stage, and the parameters used in the method.

\sout{We note that source code will be released pending publication, and a link to the repository inserted at this point.}


\sout{Moreover, $G$ does not always consist of only one connected component. Before running into the reconstruction session, we first calculate the number of connected components of $G$. Each connected component is subsequently treated independently.}

\subsection{Face identifier maintenance}
\label{sec:Face}
In order to efficiently perform the topology test described in Section~\ref{sec:topotest}, we need a data structure that allows us to quickly determine if two halfedges lie on the boundary of the same face, while also supporting fast updates as edges are inserted and faces split.

Using the Euler Tour technique~\cite{henzinger1999}, we store the sequence of halfedges of each face in a balanced binary tree with implicit keys. To query if two half-edges lie on the same face, we can simply traverse \rev{up} to the roots of the associated binary trees and compare, allowing for $O(\log |V|)$ queries. 

Storing the faces in this manner also allows for the use of standard binary tree operations, such as splits and joins. \rev{By rebalancing while performing these operations, they can be executed in $O(\log |V|)$ time while maintaining balance}. When inserting a new edge, we perform at most two splits to obtain binary trees representing the sequences of halfedges that bound the newly split face, and at most three joins to incorporate the new half-edges in these sequences. Thus, we can maintain the faces under the insertion of new edges in $O(\log |V|)$ time as well.



\subsection{Quality concern}
\label{sec:quality_concern}
To ensure the quality of the output mesh, we impose some constraints throughout the reconstruction procedure.

First, if an edge candidate results in a triangle with an angle larger than $175^\circ$ or smaller than $5^\circ$, we reject the candidate. Often, better candidate edges are \revision{connected} later, and if this does not happen, the final triangulation guarantees that the output is a triangle mesh.

When the reconstructed objects are of genus $>0$, we take an additional measure. The issue is that handle edges, which connect different faces, are not inserted during edge insertion. This, in turn, means that long edges that connect two vertices on the same side of a crack (please refer to Figure~\ref{fig:pipeline} for an illustration of a crack) may be selected. However, this could again lead to skinny triangles. Thus, for a shape that is not genus 0, we are more conservative, selecting just the first $\frac{2}{3}$ of the sorted edges from $G$ in the edge insertion stage, leaving the remaining edge insertion to the triangulation stage.

Finally, during handle connection, we only consider handle edges that connect to vertices at a corner with an angle greater than $\pi$. Again, this is to avoid skinny triangles.

\subsection{Triangulation}
\label{sec:add_face}
After the handle edges have been inserted, the binary tree which stores the association of faces to halfedges is no longer valid. In principle, we could have kept it updated, but at this point, the remaining large faces (in terms of number of edges) correspond to holes, and the remaining small faces can be triangulated efficiently using the following procedure.

The first step of triangulation is to compute the set of potential edges which could be added. We start by visiting all vertices associated with the connected handles. For each vertex, $u$, we consider all outgoing halfedges. For each outgoing halfedge, $h_1 = (u,v)$, we use the $\rho$ operator to find the next halfedge, $h_2 = (u,w)$ and then the incident vertex, $w$. Our edge candidate is now $\left\{v,w\right\}$. In most cases $\left\{v,w\right\}$ is already in $M$. Otherwise, \revision{adding the edge will "clip" a triangle from the face, and it} is pushed onto a priority queue, $Q_t$. In each iteration of the triangulation loop, we pop an edge candidate. The candidate edge is inserted if the corner of the opposite vertex (i.e. $u$) has not been split, since forming a triangle by inserting the edge would then lead to a non-manifold configuration.

If the edge is inserted, the incident vertices will be visited to search for more edge candidates. If no more candidates can be found, we traverse all vertices in $M$ that have not been visited to ensure all remaining faces are triangulated.

\subsection{Time complexity analysis}
\subsubsection{Initialization}
The initialization includes the construction of a 3D kD-tree, computation of an MST, initialization of $RS$, and sorting edges in $G$. Construction of such a kD-tree can be done in $O(|V|\log |V|)$. With the kD-tree constructed, we generate $G$ by use of $k$NN queries on the tree. In the worst case, each query takes $O(|V|)$ time to answer, but for practical purposes, the expected running time of each query is $O(\log |V|)$\cite{friedman-knn}. We can then compute the MST with an algorithm such as Prim's algorithm in $O(k|V|\log |V|)$ time. Finally, we traverse the MST and $G$ to construct $RS$ and sort the edges, where the worst case takes $O(|V|\log |V|)$.

The worst-case total time for initialization is thus $O(|V|^2)$, due to the pessimistic bounds on answering $k$NN queries. For non-adversarial input, one would expect $O(k|V|\log |V|)$ time.

\subsubsection{Edge insertion}
The running time of the edge insertion stage is dominated by the \revision{$k$NN} queries used to inspect the local geometry. Recall that we have constructed a kD-tree of the input points and that a pessimistic worst-case bound on answering these queries is $O(|V|)$. It follows then that there are at most $O(k|V|)$ edges to consider, each incurring $O(|V|)$ worst-case costs, totaling in $O(|V|^2)$ time spent. In practice, the number of edges must be much smaller, and the time to answer $k$NN queries might behave closer to $O(\log |V|)$.
\subsubsection{Handle connection}
Recall that the shortest path between the incident vertices of each handle needs to be calculated. To do this, we run a restricted Dijkstra's algorithm that stops early if the distance threshold has been reached.

Note that the number of edges in the graph is bounded from above by $O(k|V|)$ since we are searching through a subgraph of $G$. Our worst case bound for this restricted search is thus $O(k|V|\log |V|)$, attributed to the fact that we might not reach the early stopping condition before the entire graph is explored.

The candidate edges for handle insertion are obtained from $G$. There are thus $O(k|V|)$ candidates in the worst case. Therefore, we perform these restricted Dijkstra searches at most $O(k|V|)$ times, totaling in $O(k^2|V|^2\log |V|)$ time in the worst case, though this is highly unlikely to happen in practical usage.
\subsubsection{Triangulation}
To perform triangulation, a bound on the worst-case number of candidate edges is $O(k|V|)$. We construct a priority queue on these candidate edges, perform our validity check, and potentially add them to our structures. The total cost per edge is then $O(\log |V|)$, totaling in $O(k|V|\log |V|)$ time to triangulate.
\subsubsection{Total time complexity}
\begin{figure}[h]
    \centering
    \includegraphics[width=\columnwidth]{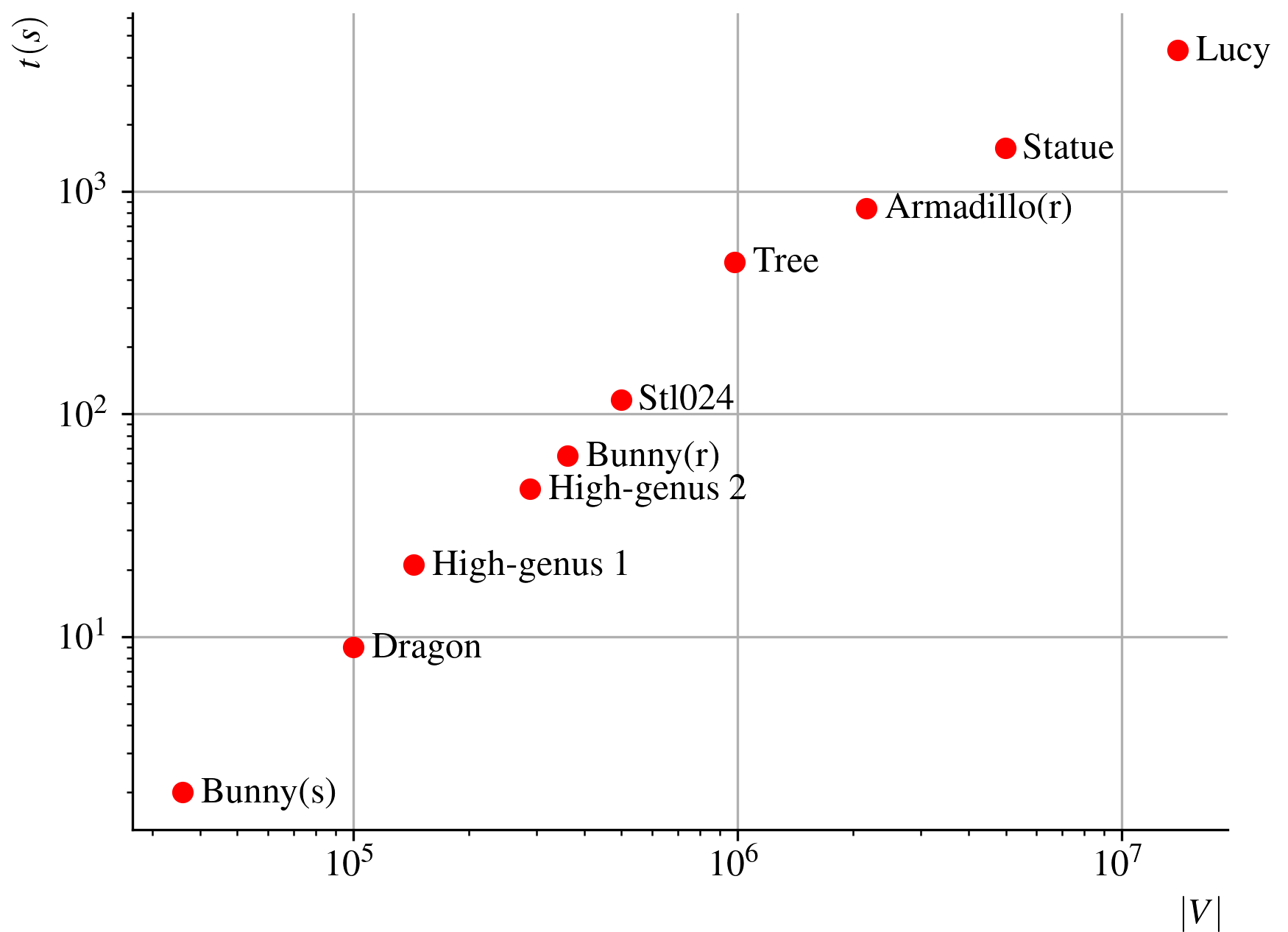}
    \caption{\revision{Time consumption on each shape in our results. Both axes are plotted in log scale, and the mesh name is annotated at the right of the data point. Notably, (s) stands for synthetic point cloud, while (r) stands for real scanning point cloud.}}
    \label{fig:time_plot}
\end{figure}

Taking the fact that the time complexity in the worst case of face identifier maintenance and triangulation is $O(\log |V|)$ and $O(k|V|\log |V|)$. Theoretically, the worst time complexity is dominated by handle connection, which is $O(k^2|V|^2\log |V|)$. In practice, however, the most time-consuming procedure is edge insertion since there are many more regular edges than handle edges even for objects of very high genus. In practice, we plot the time spent on the majority of tested shapes in Section~\ref{sec:result}, as depicted in Figure~\ref{fig:time_plot}. The results demonstrate an almost linear increase in time as the input number of vertices, $|V|$, increases.

\subsection{Parameter settings}
\label{sec:parameter}
Our method relies on several parameters. These are all discussed in Section~\ref{sec:method}\sout{ and Section~\ref{sec:implementation}}. Below, we summarize the significance and default value of the four parameters we consider user exposable.
\begin{itemize}
    \item{$k$} (default: $30$): the number of neighbors to which each point is connected in the initial graph $G$. \revision{It decides the initial connectivity of $G$. All edges connected in later stages are from this $k$NN graph.}
    \item{$r$} (default: $20$): edges longer than $r$ times the average edge length are culled. \revision{Some vertices are outliers all of whose incident edges are \rev{significantly} longer than the average edge length, and some connections are between points sampled from different surface components.}
    \item{$\theta$} (default: $60^\circ$):
    edges in $G$ for which the angle between the normals associated with the endpoints is greater than $\theta$ are removed. \revision{$\theta$ is used to reject edges whose incident vertices' normals are affected by significant noise, or edges connecting vertices from different surface components.}
    \item{$n$} (default: $50$): if the number of edges in the shortest path between the endpoints of a handle edge is less than $n$ it is rejected. \revision{Local spurious handles are eliminated by setting this threshold.}
\end{itemize}


It is reasonable to consider whether we should eschew the first three parameters and simply form a complete graph. In principle, this could be advantageous and would allow the algorithm to close holes of arbitrary size. Unfortunately, the cost in terms of both memory and run time would be prohibitive as indicated by the complexity analysis.



\begin{figure*}[p]
    \centering
    \includegraphics[width=\textwidth]{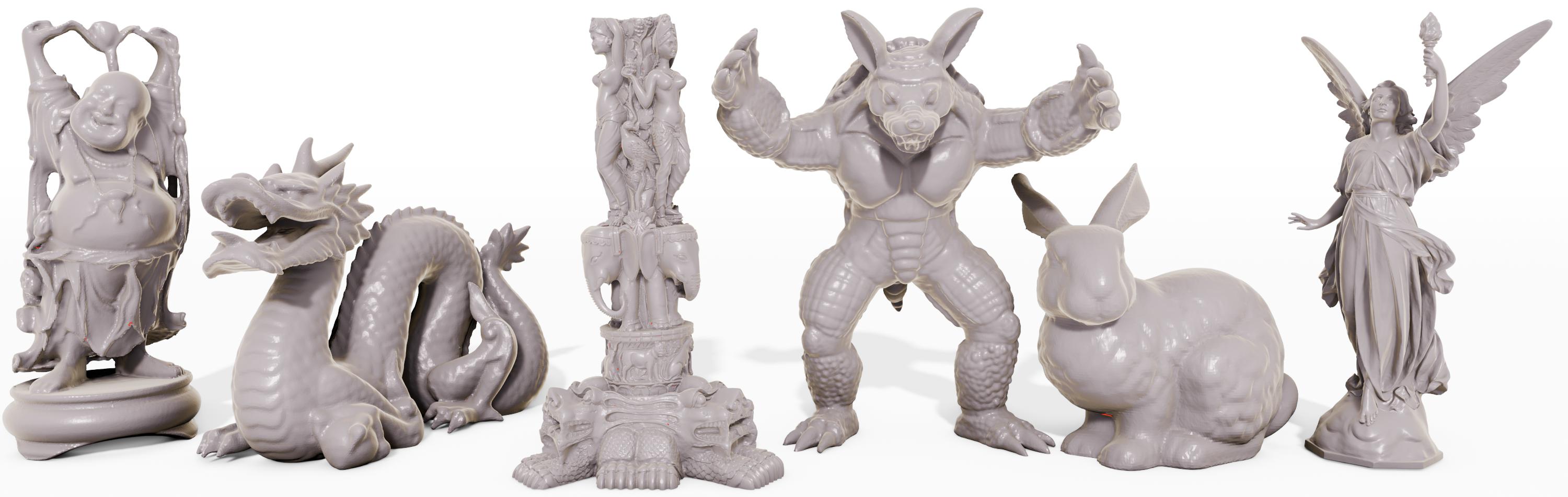}
    \includegraphics[width=\textwidth]{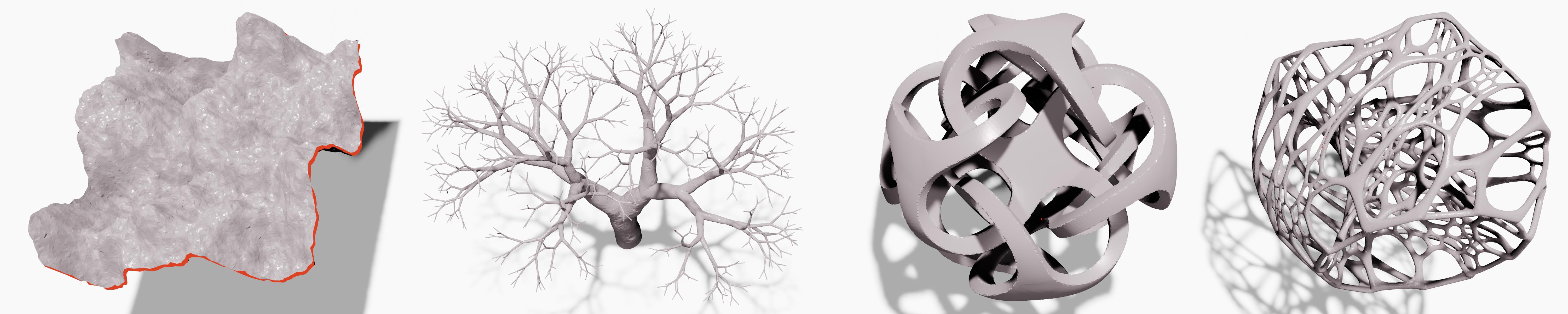}
    \includegraphics[width=\textwidth]{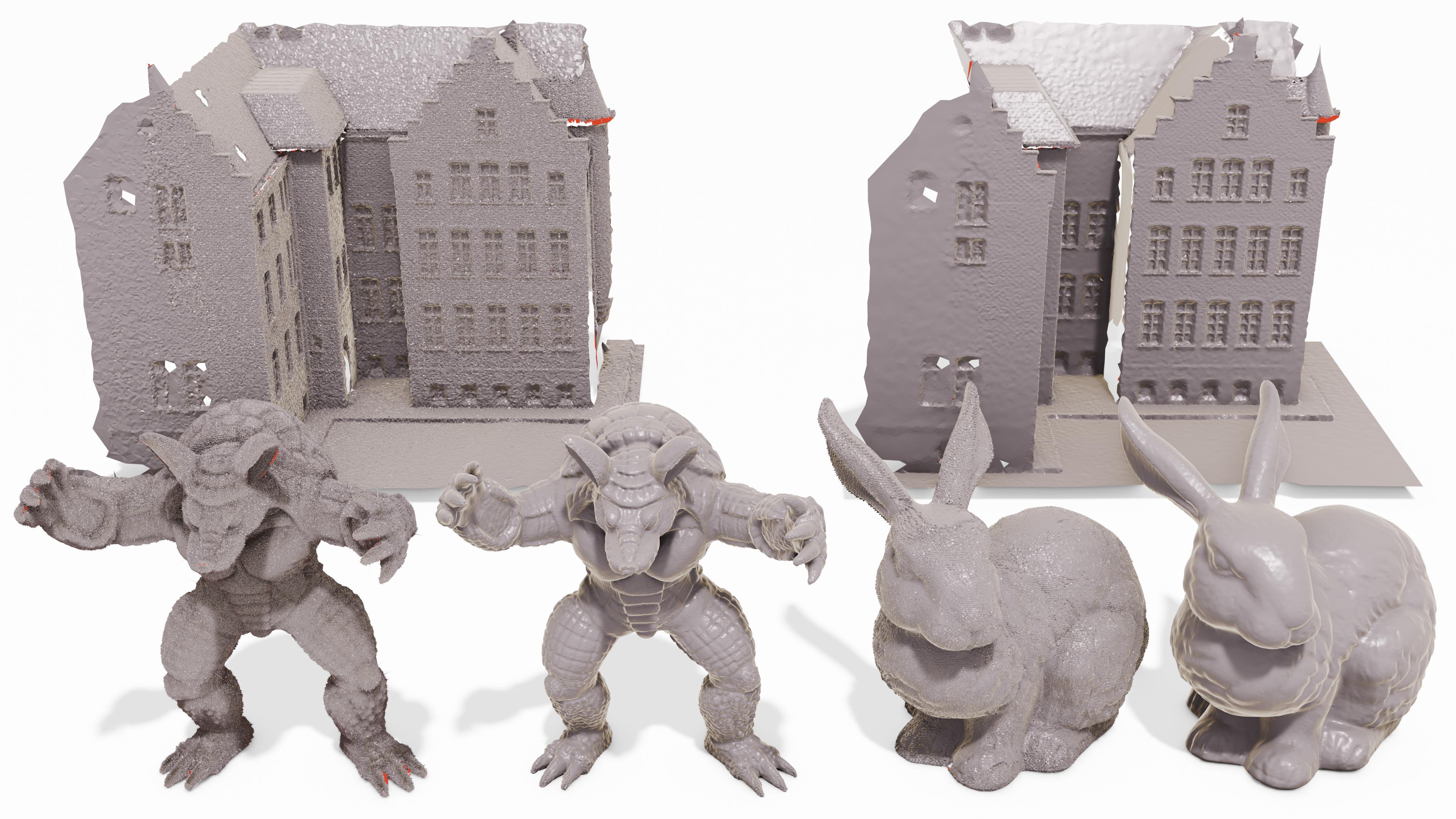}
    \caption{Renderings of models reconstructed using our method. Top image: well-known models from the Stanford 3D Scanning Repository. Middle image: several test objects are shown. The terrain (shown after hole closing) consists of two very close surfaces, the tree exhibits significant variation point density, and the two remaining objects are high-genus shapes. Bottom: we used our method to reconstruct the DTU STL0024 model as well as the Stanford Armadillo and Bunny directly from the acquired points. For each model, we see unprocessed reconstructions on the left and models after hole closing and denoising on the right. In all images, back\revision{-}facing surfaces are rendered in red to show holes in the reconstruction.
    }
    \label{fig:special_synthetic_recon}
\end{figure*} 

\section{Results}
\label{sec:result}







\begin{table*}[h]
\centering
\caption{Quantitative results on a diverse range of data. Models in the top half are from synthetic datasets while those in the bottom half are from scanning datasets. All four indicators are better when smaller. As most meshes are not originally watertight, the number of boundary edges will barely decrease to zero. To maintain a reasonable runtime, the point cloud from the DTU dataset (i.e. Stl002, Stl003, and Stl024) is reduced to approximately 500,000 points through random sampling.}
    \label{tab:quant_result_cmp}
    \resizebox{\textwidth}{!}{
    \begin{tabular}{|*{19}{c|}}
    \hline
    \multirow{2}{*}{Model Name} & \multirow{2}{*}{$|V|$} & \multicolumn{3}{c|}{Ours} & \multicolumn{3}{c|}{ScaleSpace} & \multicolumn{3}{c|}{\makecell{Co3Ne \\ w/o smoothing}} & \multicolumn{3}{c|}{\makecell{Co3Ne \\ w/ smoothing}} & \multicolumn{3}{c|}{DSE Meshing} & \multicolumn{2}{c|}{Poisson} \\
    \cline{3-19}
    & & $r_v (\%)\uparrow$ & $|E_b|\downarrow$ & $t (s)\downarrow$ & $r_v (\%)\uparrow$ & $|E_b|\downarrow$ & $t (s)\downarrow$ & $r_v (\%)\uparrow$ & $|E_b|\downarrow$ & $t (s)\downarrow$ & $r_v (\%)\uparrow$ & $|E_b|\downarrow$ & $t (s)\downarrow$ & $r_v (\%)\uparrow$ & $|E_b|\downarrow$ & $t (s)\downarrow$ & |V| & \revision{$t (s)\downarrow$}\\

    \hline

    High-genus 2 & 288,182 & 99.996 & \textbf{200} & 46 & 52.926 & 100,850 & 97 & 99.982 & 3,720 & 0.5 & 99.769 & 11,465 & \textbf{0.3} & \textbf{100} & 783 & 764 & 253,730 & \revision{7.7}\\

    Tree & 982,923 & \textbf{99.998} & \textbf{2,574} & 482 & 63.125 & 580,363 & 3,725 & 99.893 & 23,581 & \textbf{1.2} & 99.936 & 28,851 & 1.3  & - & - & - & 1,770,349 & \revision{56}\\

    Terrain & 80,000 & \textbf{100} & \textbf{650} & 12 & 68.678 & 52,475 & 25 & 99.411 & 8,469 & \textbf{0.2} & 98.994 & 9,244 & \textbf{0.2} & 99.903 & 14,402 & 224 & 299,737 & \revision{69}\\

    High-genus 1 & 143,662 & \textbf{100} & \textbf{0} & 21 & 99.965 & \textbf{0} & 19 & \textbf{100} & 338 & \textbf{0.2} & \textbf{100} & 534 & \textbf{0.2} & \textbf{100} & 106 & 391 & 336,222 & \revision{9.0}\\
    
    Mobius band & 76,387 & \textbf{100} & \textbf{0} & 10 & 26.897 & 20,656 & 186 & 99.999 & 386 & \textbf{0.1} & 98.606 & 14,291 & 0.2 & \textbf{100} & 12 & 187 & 104,756 & \revision{3.4}\\

    \hline
    \hline
    Bunny & 362,271 & \textbf{99.999} & \textbf{276} & 65 & 99.912 & 333 & 59 & 93.070 & 62,053 & 1.2 & 99.997 & 408 & \textbf{0.4} & 99.923 & 72,928 & 857 & 689,066 & \revision{19}\\
    
    Armadillo & 2,166,950 & \textbf{99.748} & \textbf{1,074} & 841 & 98.694 & 11,602 & 792 & 81.115 & 1,284,705 & 21.7 & 98.545 & 101,031 & \textbf{5.2} & - & - & - & 2,142,910 & \revision{71}\\
    
    Stl002 & 499,501 & \textbf{99.999} & \textbf{455} & \textbf{137} & 99.868 & 2,268 & 644 & 90.890 & 177,184 & 3.7 & 99.987 & 701 & \textbf{0.9} & 99.968 & 76,443 & 1101 &  559,810 & \revision{18}\\

    Stl003 & 498,768 & \textbf{99.997} & \textbf{1,879} & 121 & 97.073 & 5,244 & 714 & 90.129 & 158,215 & 3.5 & 99.916 & 2,401 & \textbf{0.6} & 99.904 & 106,544 & 1074 & 867,621 & \revision{27}\\
    
    Stl024 & 500,000 & \textbf{99.999} & \textbf{1,387} & 116 & 98.316 & 2,171 & 831 & 94.914 & 101,693 & 2.8 & 99.971 & 2,853 & \textbf{0.4} & 99.977 & 52,972 & 1153 & 477,303 & \revision{15}\\

    \hline
    \end{tabular}
    }
\end{table*}

To evaluate our method extensively, we carried out experiments on two different types of point clouds, namely synthetic data, i.e. data sampled from existing surfaces, and real scanning data. 

The synthetic data sets are typically not noisy, while real scans, on the other hand, are very dense since they tend to be composed of several combined sub-scans. This makes it easy to estimate a relatively smooth normal field despite this type of data being affected by significant amounts of noise.

We will compare our reconstructions to several state-of-the-art combinatorial reconstruction methods as well as screened Poisson reconstruction. All experiments, except those that involve DSE Meshing, were conducted on a laptop with a CPU 13th Gen Intel(R) Core(TM) i7-1365U @ 1.8GHz and 32 GB RAM. As DSE Meshing requires GPU, it is tested on HPC with GPU Tesla V100-PCIE-16GB, CPU Intel(R) Xeon(R) CPU E5-2660 v3 @ 2.60GHz, and 128 GB RAM.
\subsection{Synthetic datasets}
\sout{In this section, we present the results of reconstruction from synthetic point clouds.}
\subsubsection{Stanford 3D Scanning Repository}
For the models from the Stanford Scanning Repository, we use the vertex positions and normals from the reconstructions \cite{curless1996volumetric} provided by Stanford Computer Graphics Laboratory.

Figure \ref{fig:special_synthetic_recon} (top) shows the reconstructed meshes. Table \ref{tab:stanford_statistics} contains practical information about the reconstructions, including the number of input vertices ($|V|$), running time ($t(s)$), and the number of output faces ($|F|$).

\begin{table}[]
\centering
\caption{Practical statistics of reconstruction on Stanford Repository. }
\label{tab:stanford_statistics}
    \begin{tabular}{|c|c|c|c|}
    \hline
    Model Name & $|V|$ & $|F|$ & $t(s)$ \\
    \hline
    Bunny & 35,947 & 71,816 & 2 \\
    Dragon & 100,240 & 200,268 & 9 \\
    Happy Buddha & 144,625 & 289,104 & 15 \\
    Armadillo & 172,967 & 345,928 & 15 \\
    Thai Statue & 4,999,518 & 9,994,987 & 1,562 \\
    Lucy & 14,025,736 & 28,051,203 & 4,301 \\
    \hline
    \end{tabular}
\end{table}

\subsubsection{Manifold ShapeNet}

\begin{figure}[h]
    \centering
    \includegraphics[width=\columnwidth]{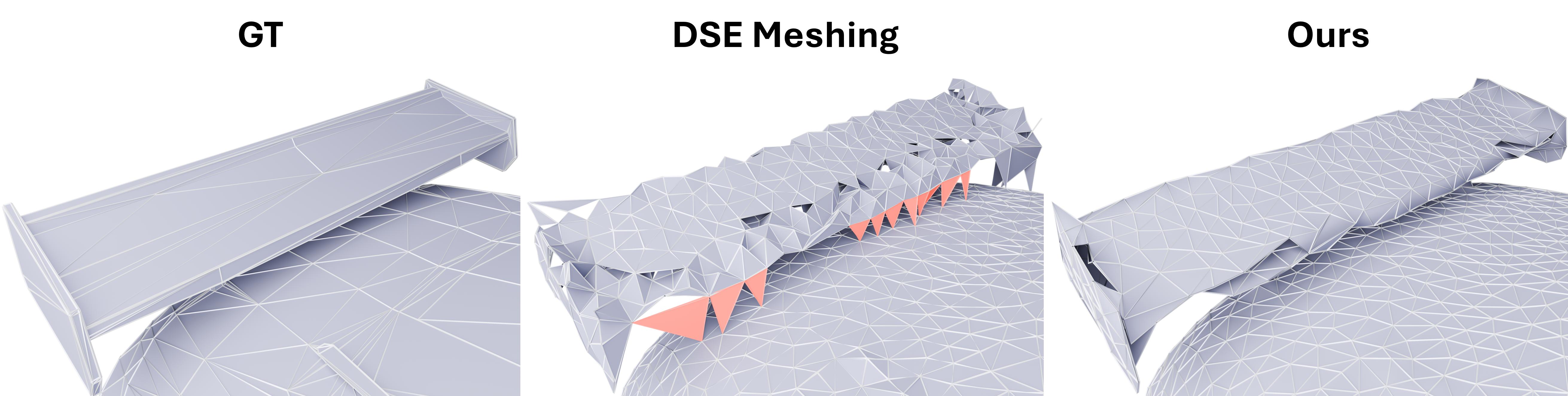}
    \caption{Qualitative result on Manifold ShapeNet. Non-manifold triangles are rendered red.}
    \label{fig:shapenet_comp}
\end{figure}

ShapeNet \cite{chang2015shapenet} is a popular dataset of 3D CAD models of objects. As many recent works use it as a main comparison dataset, we also carried out an experiment on it. However, since most meshes in this dataset are made by human designers, a large number of them are non-manifold. To perform reconstruction with reasonable input, we use the processed ShapeNet by Chu et al. \shortcite{chu2019repairing}. We directly download the data of \revision{a} processed subset of ShapeNet from \rev{\href{https://github.com/lei65537/Visual_Driven_Mesh_Repair}{this repo}}. Specifically, we use meshes after optimization and hidden patch removal, i.e. meshes with a suffix of \textbf{\_HR}. There are 1,572 shapes in total, a quantitative metric of Chamfer Distance (CD) is applied to evaluate the performance. \revision{To calculate the CD score, we don't directly use the mesh vertices. Instead, the same number of points are resampled on the reconstructed mesh.\sout{Limited by the resources,} Due to resource limitations,} we only compare with DSE Meshing in this experiment. To fit the best number of input points of DSE Meshing, we use Poisson disk sampling \cite{corsini2012efficient} to sample 10k points with MeshLab \cite{meshlab} from the ground truth (GT) meshes. It turns out our method has better control over the topology, thus generating better reconstruction. A qualitative comparison is exhibited in Figure~\ref{fig:shapenet_comp}. \revision{Quantitatively}, our method reaches an average CD score of $\textbf{0.00836}$ over the 1.5k shapes, which is better than the CD score of $0.00846$ by DSE Meshing.

\subsubsection{Special data}
\label{sec:synthetic_comp}
To assess the robustness of our method, we conducted experiments on a diverse collection of synthetic data sets exhibiting special properties. Two high-genus shapes from the Thingi10k dataset \cite{Thingi10K} were chosen to serve as a topological challenge. Additionally, a tree generated \revision{using an} L-system \cite{prusinkiewicz1996systems} \revision{\sout{method}} was employed to test our method's robustness in handling input with significant density variation. Moreover, the inclusion of a terrain (with two sheets) and \revision{a} M\"{o}bius band \revision{served} to evaluate the performance on inputs with extremely close sheets. The results of these experiments are shown in Figure \ref{fig:special_synthetic_recon} (middle).

\begin{figure}[h]
    \centering
    \includegraphics[width=0.8\columnwidth]{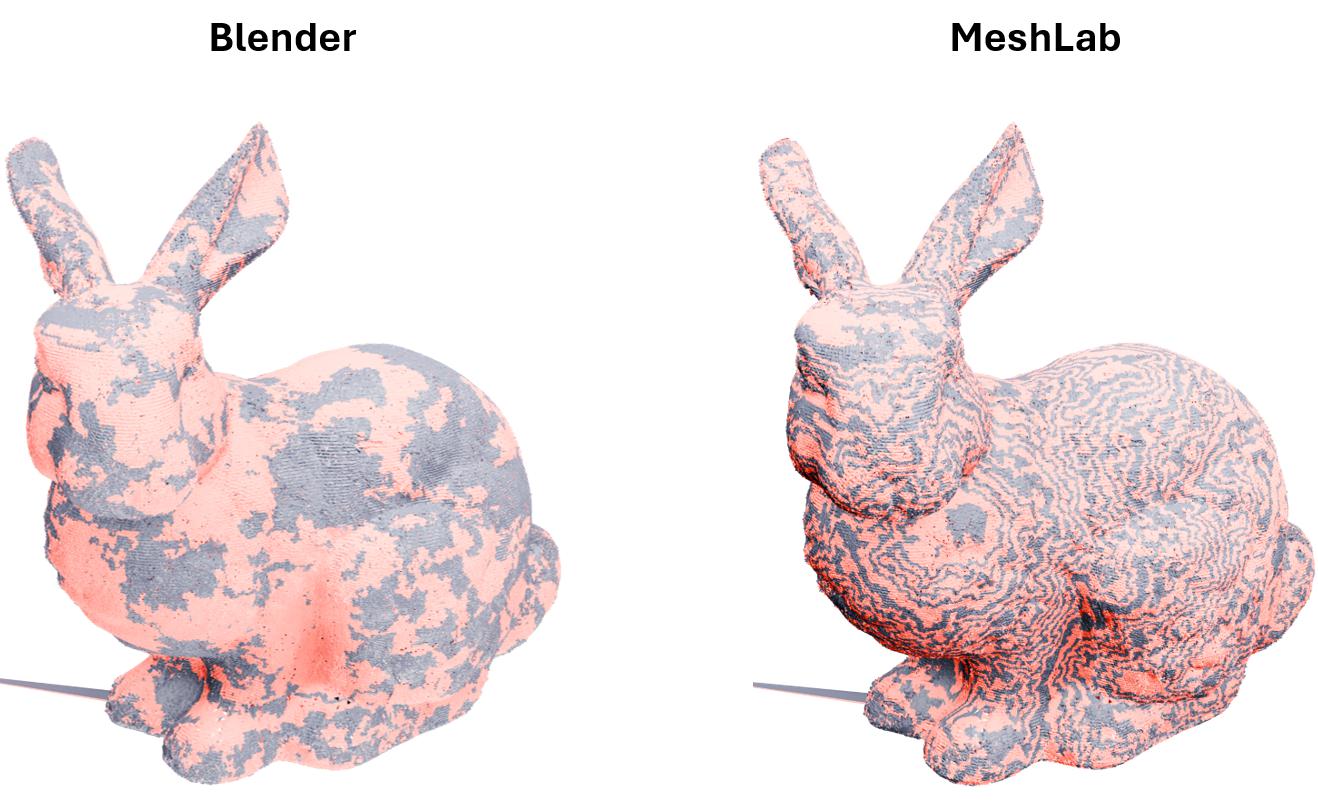}
    \caption{Reconstructed mesh of raw bunny by DSE Meshing. Both software cannot recalculate the expected face orientation.}
    \label{fig:reorient}
\end{figure}

Comparisons are conducted with three state-of-the-art combinatorial reconstruction methods, specifically Co3Ne \cite{boltcheva2017surface} (Delaunay-based), ScaleSpace \cite{Digne2011} (Ball-Pivoting Method), and DSE meshing \cite{rakotosaona2021learning} (Machine-Learning-based), as well as the popular Screened Poisson Reconstruction (SPR) \cite{kazhdan2013screened}. For the three combinatorial methods, we use the source code made available by the authors. For SPR, we use the implementation from MeshLab \cite{meshlab} with the default parameters. Also, in the interest of fairness, we generate output meshes with at least as many points as in the input data when doing SPR reconstruction. Co3Ne provides the option to apply smoothing to the input point cloud, and we include both Co3Ne with two iterations of smoothing and Co3Ne without smoothing in the comparison. 

\begin{figure*}
    \centering
    \includegraphics[width=\textwidth]{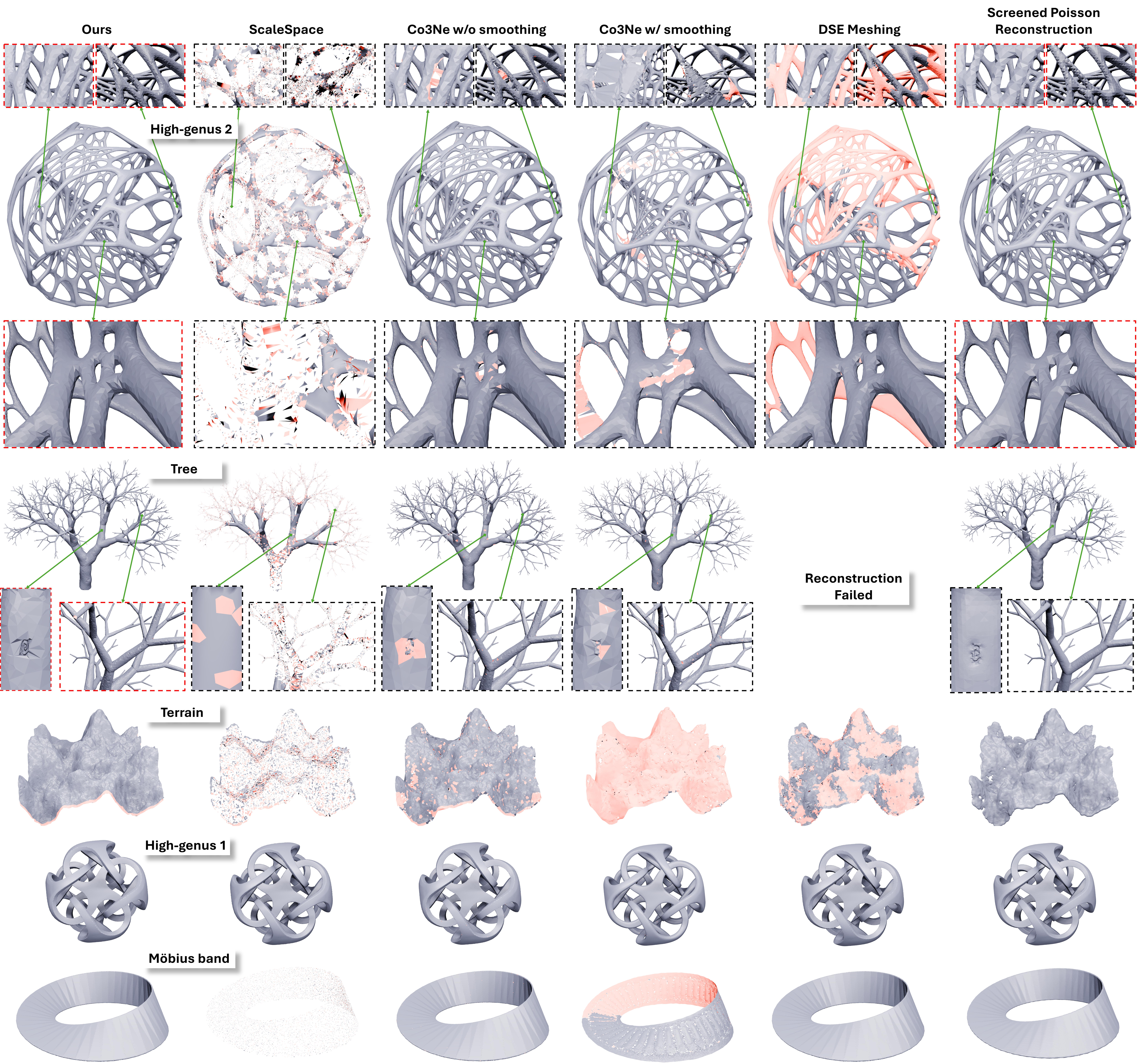}
    \caption{Visualization of the reconstruction results from six different methods on special synthetic data. A detailed image enclosed by red dashed lines indicates the best reconstruction outcome.}
    \label{fig:synthetic_cmp}
\end{figure*}

The results of the comparisons are shown in Figure~\ref{fig:synthetic_cmp}. Note that we render the back faces in red to highlight holes in the reconstruction. However, the output of DSE Meshing exhibits incorrectly oriented triangles, so in this case red means incorrect orientation. We tried built-in methods from Blender\cite{blender} and MeshLab\cite{meshlab} to recalculate the orientation. However, neither application was able to correctly orient the result as illustrated in Figure~\ref{fig:reorient}. This indicates that DSE produces topologically flawed meshes that cannot be correctly oriented. For the results shown in Figure~\ref{fig:synthetic_cmp}, we have arbitrarily used Blender to reorient the meshes.

Below, we provide our observations for each model regarding the relative merits of the various methods.
\begin{itemize}
    \item \textbf{High-genus 2}: On this shape, which is created by a human designer, only our method and Screened Poisson managed to accurately reconstruct the correct topology in the majority of locations (note that in places the source shape is non-manifold). The other three methods either created spurious handles or failed to reconstruct the shape.    
    \item \textbf{Tree}: This shape has multiple challenging properties. Firstly, the point density varies significantly: there are small patches with very high point density on the trunks, and the density is also much higher on the branches relative to the trunks. Secondly, the branches on the top of the tree are thin structures. In addition, there is noise along the normal direction of the surface. From the comparison, it turns out that only our method can accurately reconstruct both the textures and the general shape of the tree. Notably, DSE Meshing fails in the final triangle selection phase and does not produce an output mesh.

    \item \textbf{Terrain}: This shape is again a thin structure with two layers pointing in opposite directions. Our method managed to reconstruct both surfaces without leaving a hole while the other methods either left holes (Co3Ne, DSE Meshing) or connected the opposite layers (Screened Poisson). Methods with smoothing (ScaleSpace, Co3Ne) will fail the reconstruction because the pre-processing mixes points from both layers. Since the neighborhood we choose to fit the tangent plane is filtered by the normal direction, our method does not suffer from the issue above.

    \item \textbf{High-genus 1} and \textbf{M\"{o}bius band}: Broadly speaking, most methods reconstruct these shapes well. That said, the methods that smooth the input points, i.e. ScaleSpace and Co3Ne, do not perform well on the M\"{o}bius band. Moreover, local topological issues arose and will be discussed below.
\end{itemize}

Quantitative statistics for the synthetic data comparisons are provided in Table~\ref{tab:quant_result_cmp}. 
Three metrics are considered including vertex reference ratio ($r_v (\%)$), the number of boundary edges ($|E_b|$), and efficiency ($t (s)$). The definition of vertex reference ratio is as follows:
\begin{equation} 
    r_v = \frac{|V_o|}{|V_i|} \enspace ,
\end{equation}
where $|V_o|$ is the number of vertices that are referenced in the final output mesh, and $|V_i|$ is the number of points as input (A referenced vertex means that it is included in at least one triangle of the mesh). This metric is used to evaluate to what extent the input information is preserved.

Because Screened Poisson Reconstruction represents a different methodological approach, we only report the number of output vertices \revision{and the time consumption} in meshes generated using SPR.

\begin{figure}
    \centering
    \includegraphics[width=\columnwidth]{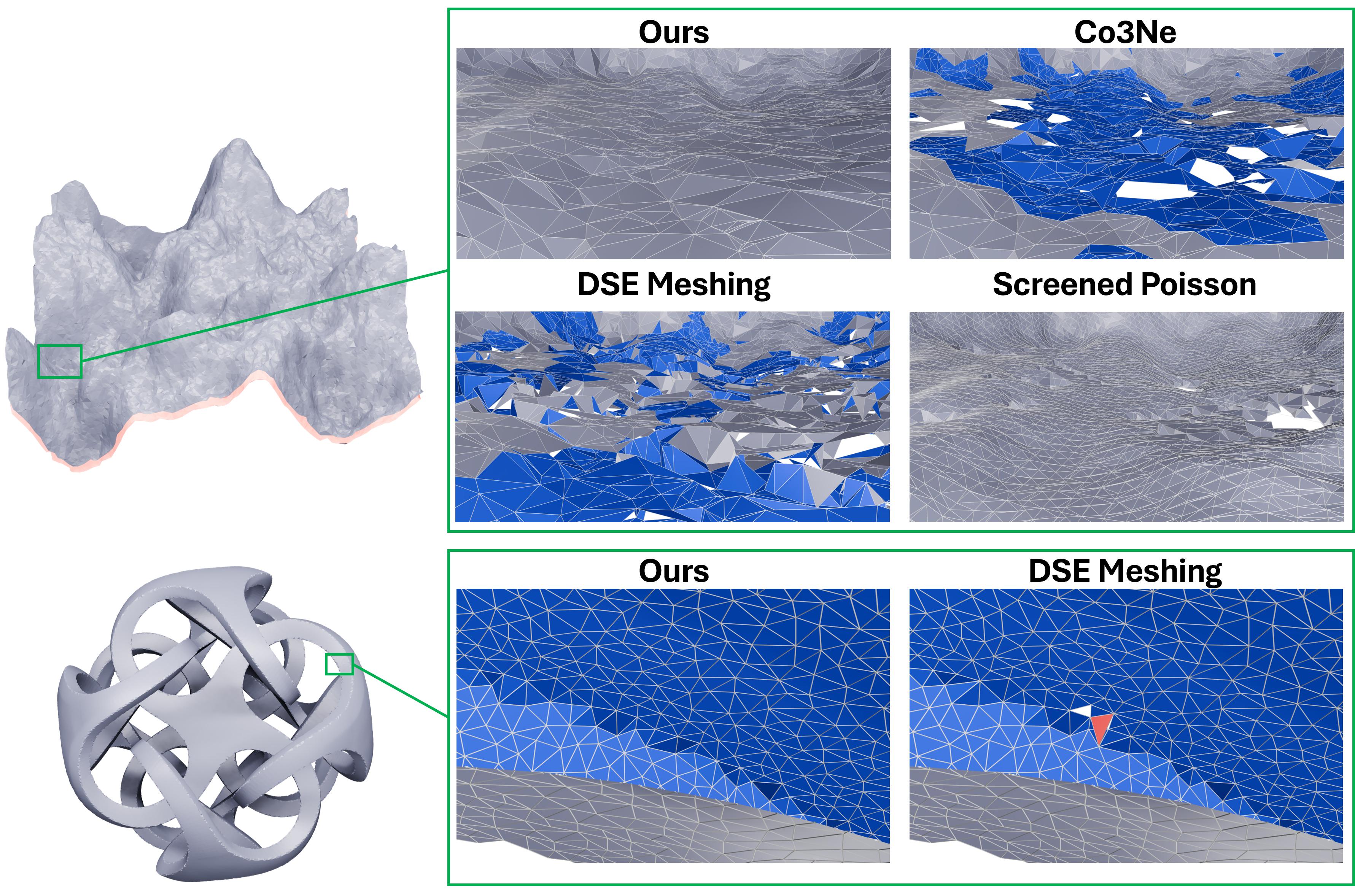}
    \caption{Examples illustrating that Co3Ne, DSE, and SPR do fail locally on the two otherwise well-reconstructed meshes. The backfaces are here displayed in blue. In the case of Terrain, the three methods Co3Ne, DSE Meshing, and Screened Poisson connect the opposite layers. In High-genus 1, DSE Meshing generates spurious handles underneath the surface at multiple locations. A handle triangle is here shown \revision{in} red.}
    \label{fig:local_error}
\end{figure}

Although the methods we compare against seem to perform well on Terrain, High-genus 1, and M\"{o}bius band, they do produce either incorrectly oriented parts or spurious handles in some places. We show these local issues in Figure~\ref{fig:local_error}.

\begin{figure}
    \centering
    \includegraphics[width=\columnwidth]{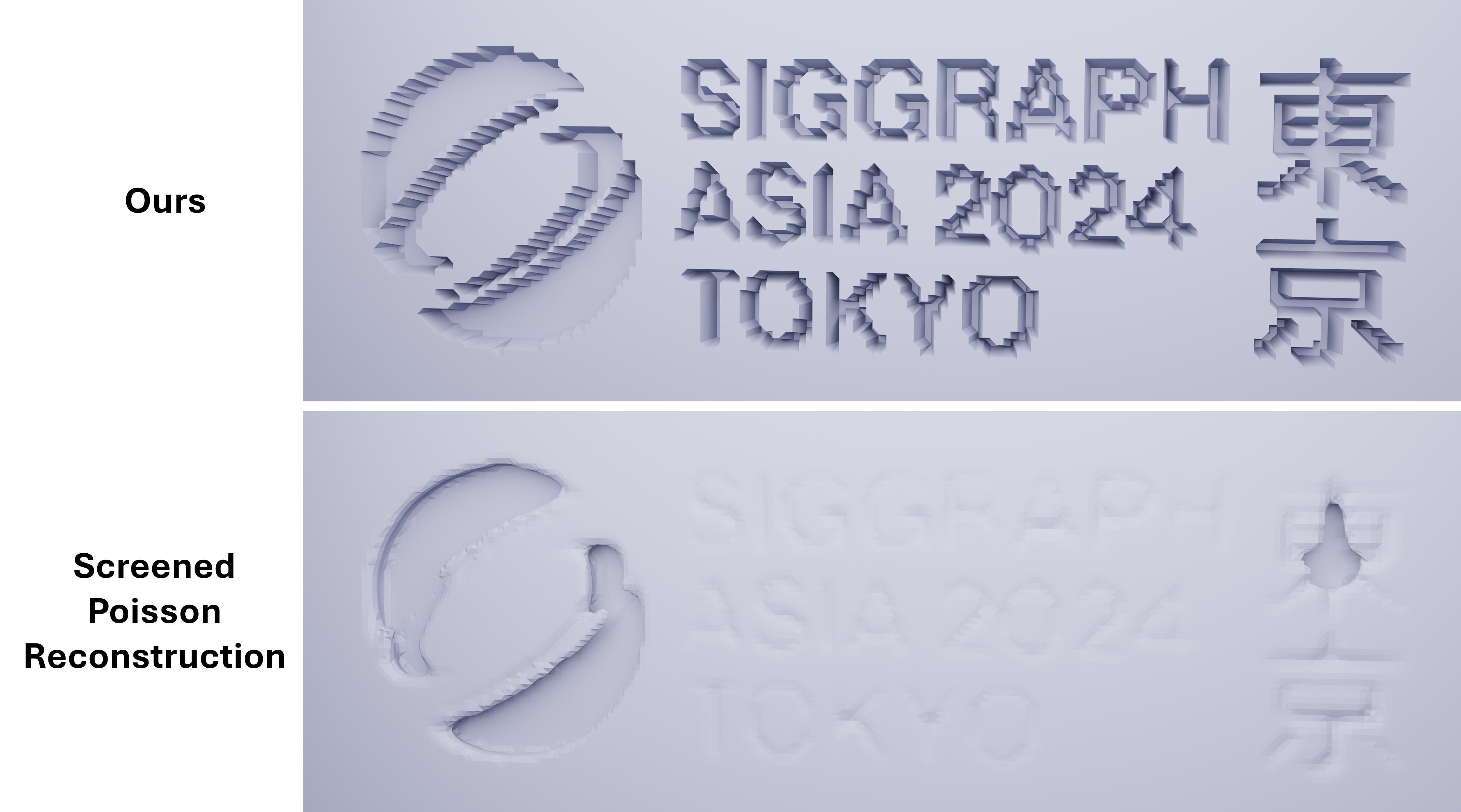}
    \caption{A synthetic example for the reconstruction of a surface with informative texture.}
    \label{fig:logo}
\end{figure}

\revision{As mentioned, volumetric methods have many advantages in terms of robustness and speed, but compared to combinatorial methods they sometimes either miss or blur subtle features. To illustrate this issue, we generated a point set with a geometric texture showing the conference logo by keeping the pixel position as the x and y coordinates while converting the pixel intensity to depth as the z coordinate. We show a comparison between our method and SPR in Figure~\ref{fig:logo}. Even with the tree depth set to 15, SPR cannot reproduce the geometry accurately. While this is a synthetic example, it seems to show a scenario where combinatorial methods are preferable.} 

\subsection{Scanning datasets}
In this section, we present the results of our method applied directly to optically acquired scans.

For these experiments, we use the projection distance as discussed in Section~\ref{sec:projection_dist}. The experiments are performed on models from well-known public datasets: the Stanford 3D Scanning Repository, the DTU Robot Image Data Set (randomly downsampled to around 500,000 points) \cite{jensen2014large}, Tanks and Temple\cite{Knapitsch2017},  and the S3DIS dataset\cite{armeni20163d} with indoor scenes.
The same comparisons are performed as for the synthetic data.

\begin{figure}[h!]
    \centering
    \includegraphics[width=\columnwidth]{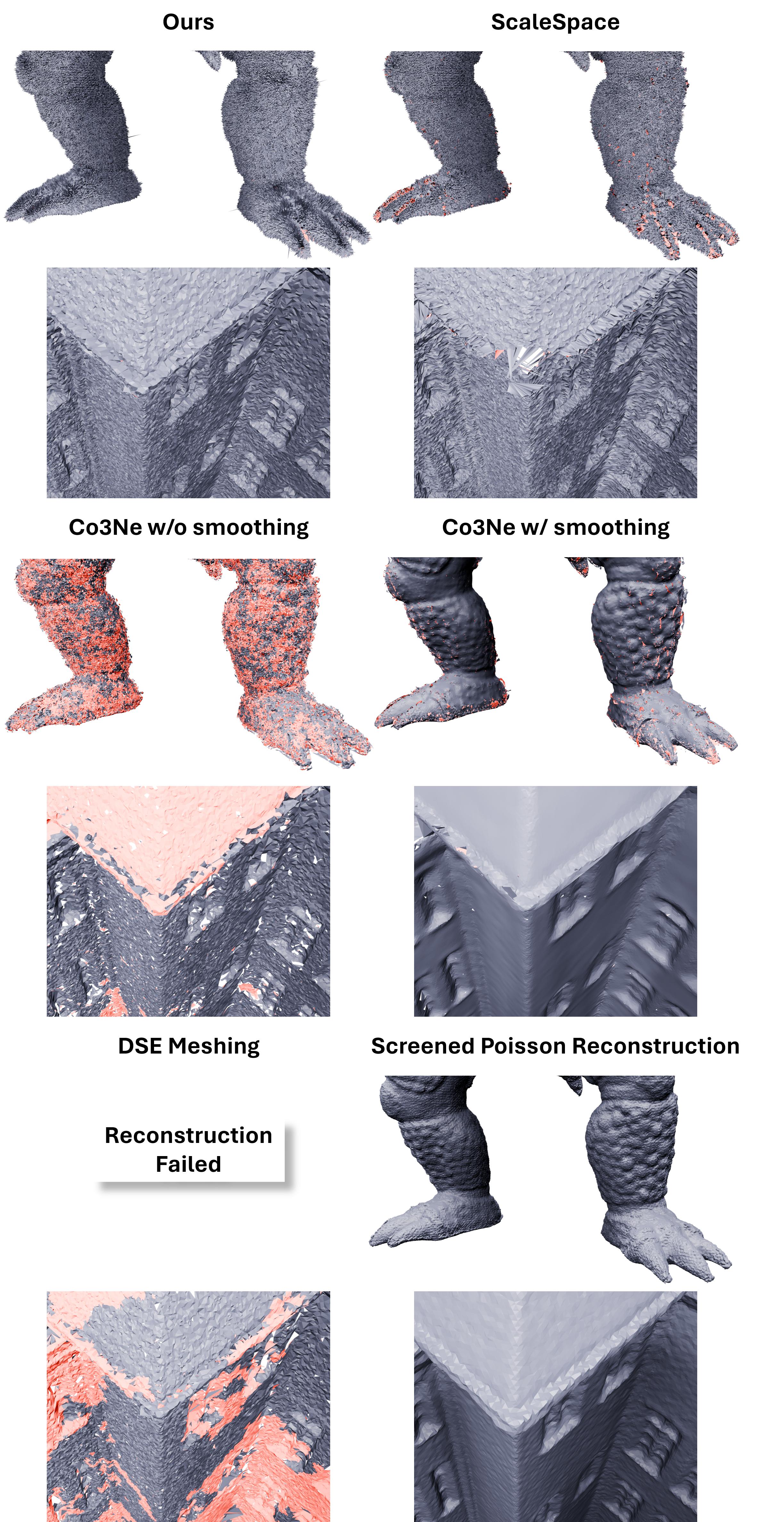}
    \caption{Details on noisy data reconstruction. The first row is the feet of Armadillo on the front side. The second row is a corner of the roof from Stl024.}
    \label{fig:scanning_detail}
\end{figure}

\begin{figure}[]
    \centering
    \includegraphics[width=\columnwidth]{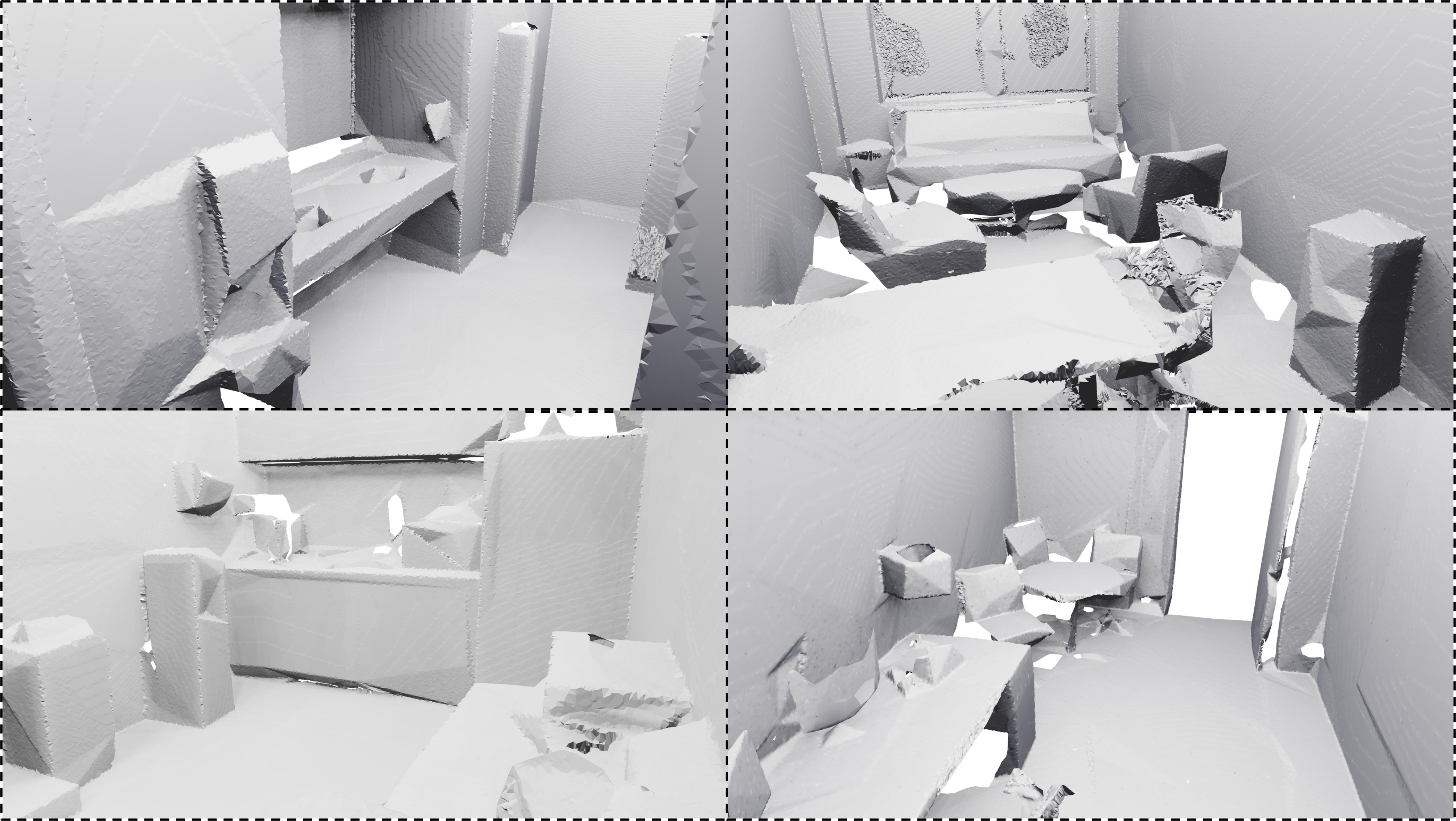}
    \caption{The reconstructed indoor scenes of four different rooms from S3DIS\cite{armeni20163d}. Most of the area, e.g. walls, chairs, and tables are well reconstructed.}
    \label{fig:indoor_scene}
\end{figure}

\begin{figure*}[]
    \centering
    \includegraphics[width=0.9\textwidth]{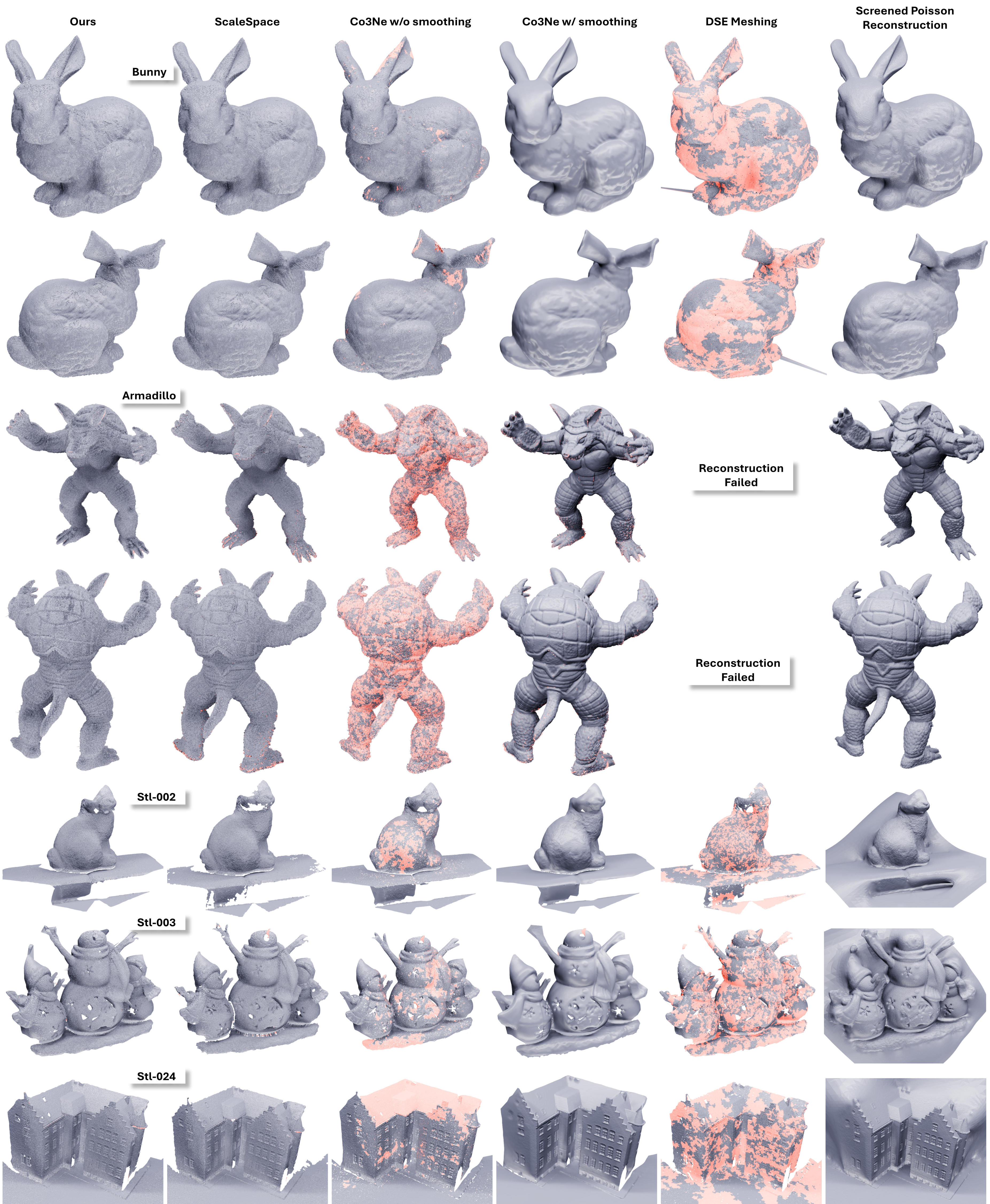}
    \caption{This is the visualization of the reconstruction results by five different methods. It can be noticed that ours, ScaleSpace, Co3Ne w/ smoothing, and Screened Poisson successfully reconstruct reasonable meshes of the input point cloud. Comparing ours and ScaleSpace, our method can reconstruct a more complete mesh with a smaller number of holes while ScaleSpace is restricted by the parameter of the Ball-Pivoting algorithm.}
    \label{fig:scanning_data}
\end{figure*}

\begin{figure*}[]
    \centering
    \includegraphics[width=0.9\textwidth]{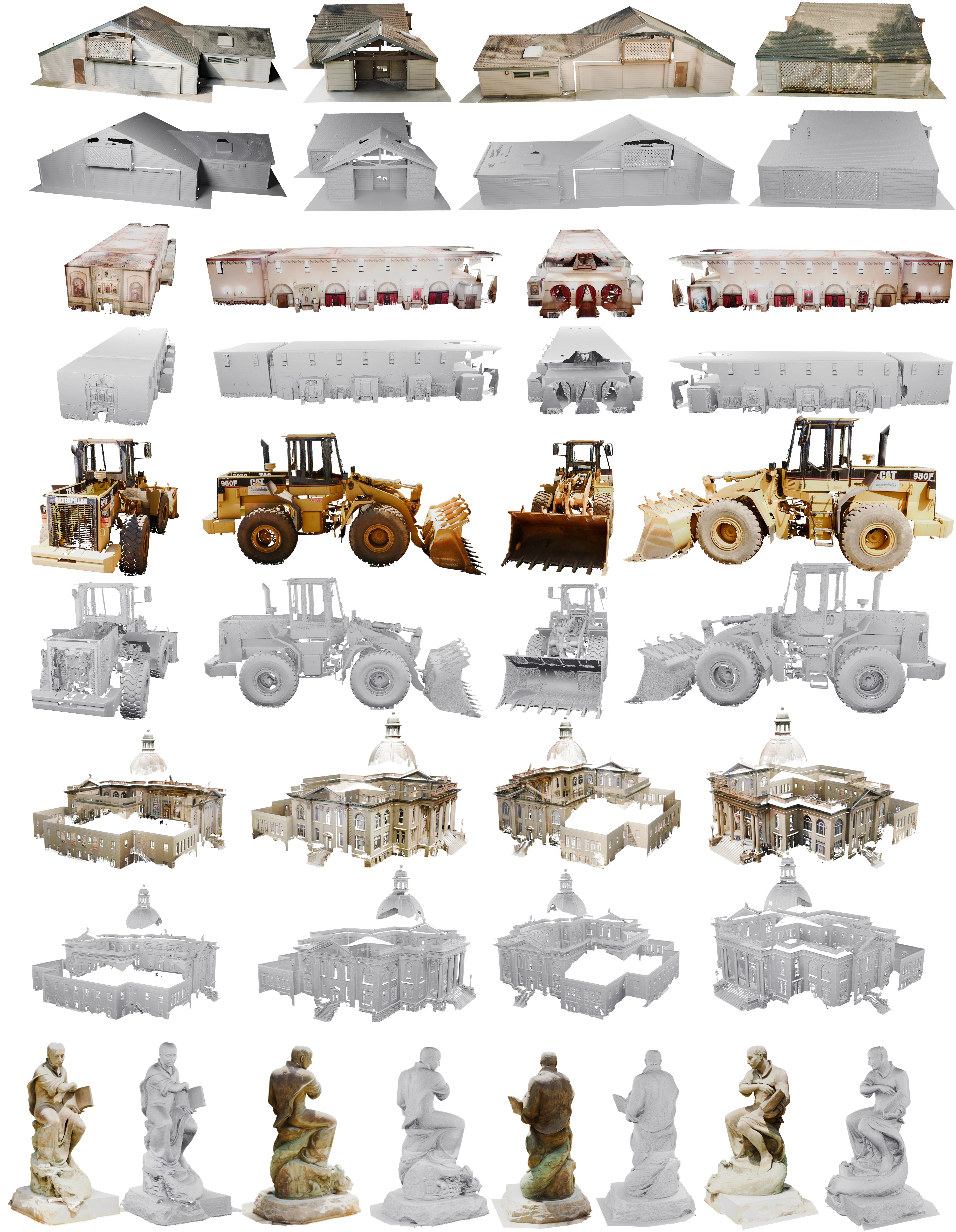}
    \caption{Partial reconstruction results of dataset Tanks and Temple \cite{Knapitsch2017}. Figures with color are the rendering of the input point cloud while those in white are the reconstruction results from our method.}
    \label{fig:tank}
\end{figure*}

Figure~\ref{fig:scanning_data} shows a comparison of our method against the four benchmark methods. Specifically, DSE Meshing failed to reconstruct the Armadillo. Restricted by the size of the figure, more details are exhibited in Figure~\ref{fig:scanning_detail}. Figure~\ref{fig:indoor_scene} and Figure~\ref{fig:tank} show the results on S3DIS and Tanks and Temple respectively. Table~\ref{tab:quant_result_cmp} contains the quantitative comparison. Figure~\ref{fig:special_synthetic_recon} (bottom) illustrates three of the scanned models before and after denoising using bilateral normal filtering \cite{Zheng2011}.

\begin{figure*}[h]
  \centering
  \includegraphics[width=\textwidth]{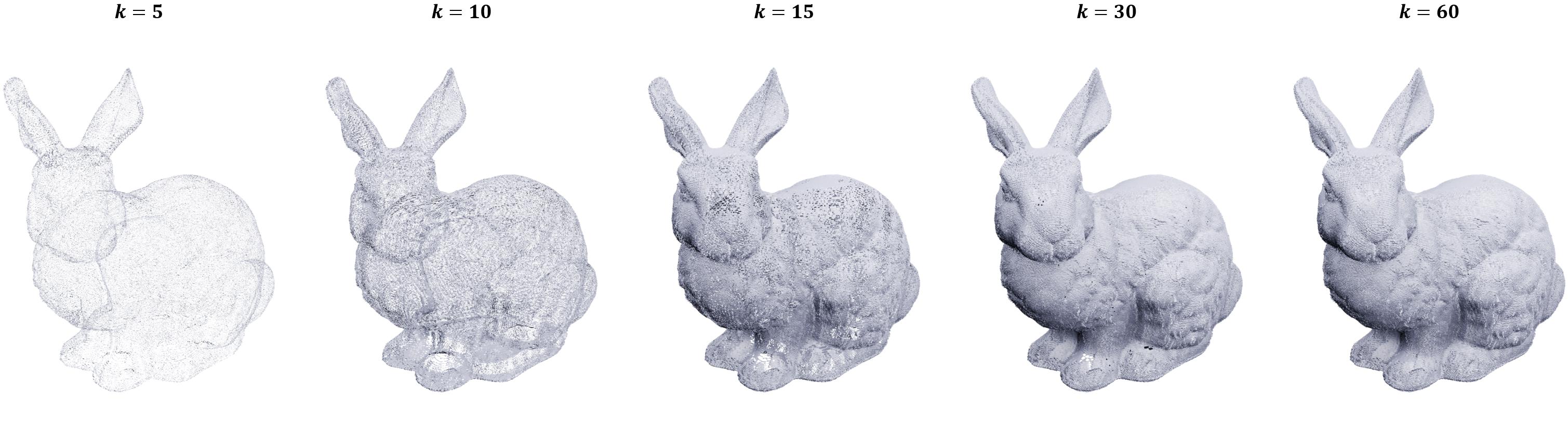}
  \caption{Qualitative result when increasing parameter $k$. \revision{The parameter $k$ determines the number of neighbors each vertex connects to in the initial graph $G$.} The larger $k$ is, the more possible connections can be expected in the reconstructed result.}
  \label{fig:k}
\end{figure*}

\begin{figure*}[h]
  \centering
  \includegraphics[width=\textwidth]{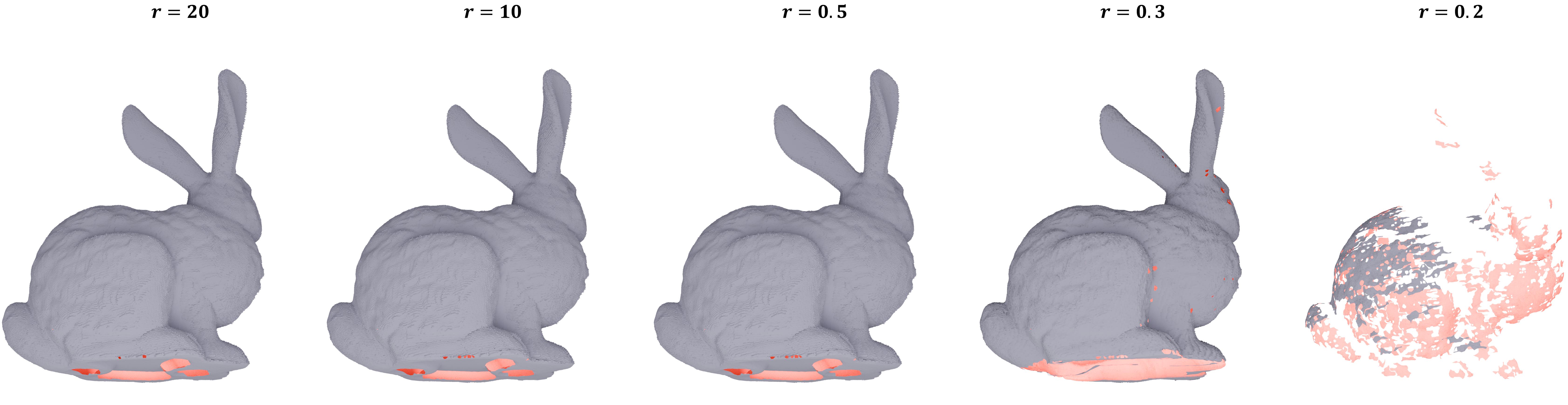}
  \caption{\revision{Qualitative result when decreasing parameter $r$. This parameter ensures that edges longer than $r$ times the average length of connections in $G$ are rejected as outliers. $k$ is set to be 50. Back faces are rendered in red.}}
  \label{fig:r}
\end{figure*}

\subsection{Parameter experiment}
As discussed in Section~\ref{sec:parameter}, we conducted an experiment to provide a clearer insight into the parameter $k$ by varying its values. Figure~\ref{fig:k} showcases the reconstructed results on the raw data of the Stanford Bunny with different values of $k$.  Table~\ref{tab:k} presents corresponding statistics for running time ($t(s)$) and peak memory usage ($m(GB)$). Unsurprisingly, resource utilization increases with $k$ while we see an increase in the number of holes for small $k$. \revision{Similarly, we carried out another parameter experiment for $r$. Figure~\ref{fig:r} demonstrates how $r$ affects the reconstruction result. Since it is used to filter out outlier connections, a small $r$ will result in a large number of edges rejected. Notably, the average edge length mentioned in Section~\ref{sec:parameter} is related to $k$, because a larger $k$ will introduce longer connections thus increasing the average edge length.}

\begin{table}[h]
\centering
\caption{Statistics for reconstruction with a range of $k$ value. Resource usage clearly increases with $k$ although the difference is slight for small $k$. However, for larger $k$ we see a significant increase in both memory consumption and time. This is largely due to the initialization where the larger graph is more costly.}
\label{tab:k}
    \begin{tabular}{|c|c|c|c|c|c|}
    \hline
    $k$ & 5 & 10 & 15 & 30 & 60 \\
    \hline
    $t(s)$ & 15 & 17 & 24 & 33 & 59 \\
    \hline
    $m(GB)$ & 0.58 & 0.70 & 0.77 & 1.1 & 1.5\\
    \hline
    \end{tabular}
\end{table}

\subsection{Explicit topology control}
\label{sec:topology-control}
\begin{figure*}[h]
    \centering
    \includegraphics[width=\textwidth]{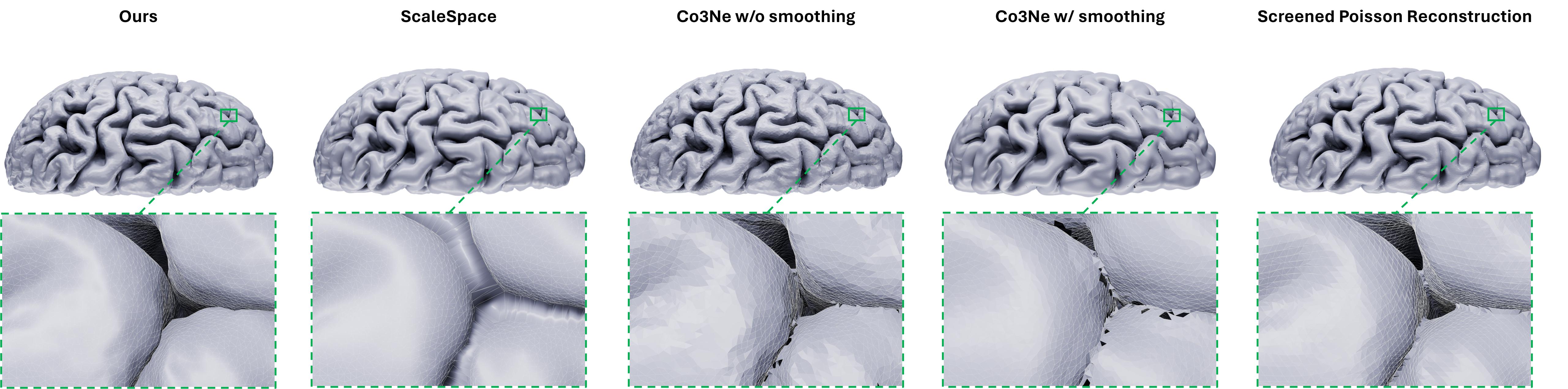}
    \caption{Reconstruction comparisons using a synthetic cortical surface point cloud. Our method successfully generated a genus-0 mesh, whereas Co3Ne retained only one side of the compressed opposite points, and Scalespace formed bridges across the gap.}
    \label{fig:topology_control}
\end{figure*}

In this section, we conduct an experiment to emphasize a key strength of our method. The explicit control we have over the number of handles being connected allows us to control the output topology. A noteworthy and extensively studied shape is the cortical surface which always has genus 0. By specifying the anticipated genus within the program, our method can omit adding handles during the reconstruction process. In contrast, other methods consistently struggle to achieve this level of topology control, as illustrated in Figure \ref{fig:topology_control}. The experimented mesh is obtained from \href{https://brainder.org/research/brain-for-blender/}{Brainder}, created by Anderson Winkler.

\subsection{Stress test}

To investigate the robustness of our method against various common defects in input point clouds, several experiments were conducted. Common defects include noise in the point position, variations in \revision{the} normal direction, and misalignment. All experiments in this section are carried out on the raw data of \revision{the} Stanford Bunny, which is, of course, noisy even before the synthetic noise is added.
\subsubsection{Noise - point position}
In an attempt to disentangle the influence of noise on point positions and normal directions, we compute the normal direction before the noise is added.
To make the noise independent of the mesh scale, the noise vector is defined as:
\begin{equation} 
    \mathbf{v_n} = A \cdot \overline{|e|} \cdot G(\mu,\sigma) \cdot \mathbf{v} \enspace ,
\end{equation}
where $\overline{|e|}$ is the average edge length of the reconstructed triangle mesh without noise. $A$ is the amplitude, $G(\mu,\sigma)$ gives a length following Gaussian distribution (where $\mu=0,\sigma=1$), and $\mathbf{v}$ gives a random direction in 3D space.

\begin{figure*}[h]
    \centering
    \includegraphics[width=\textwidth]{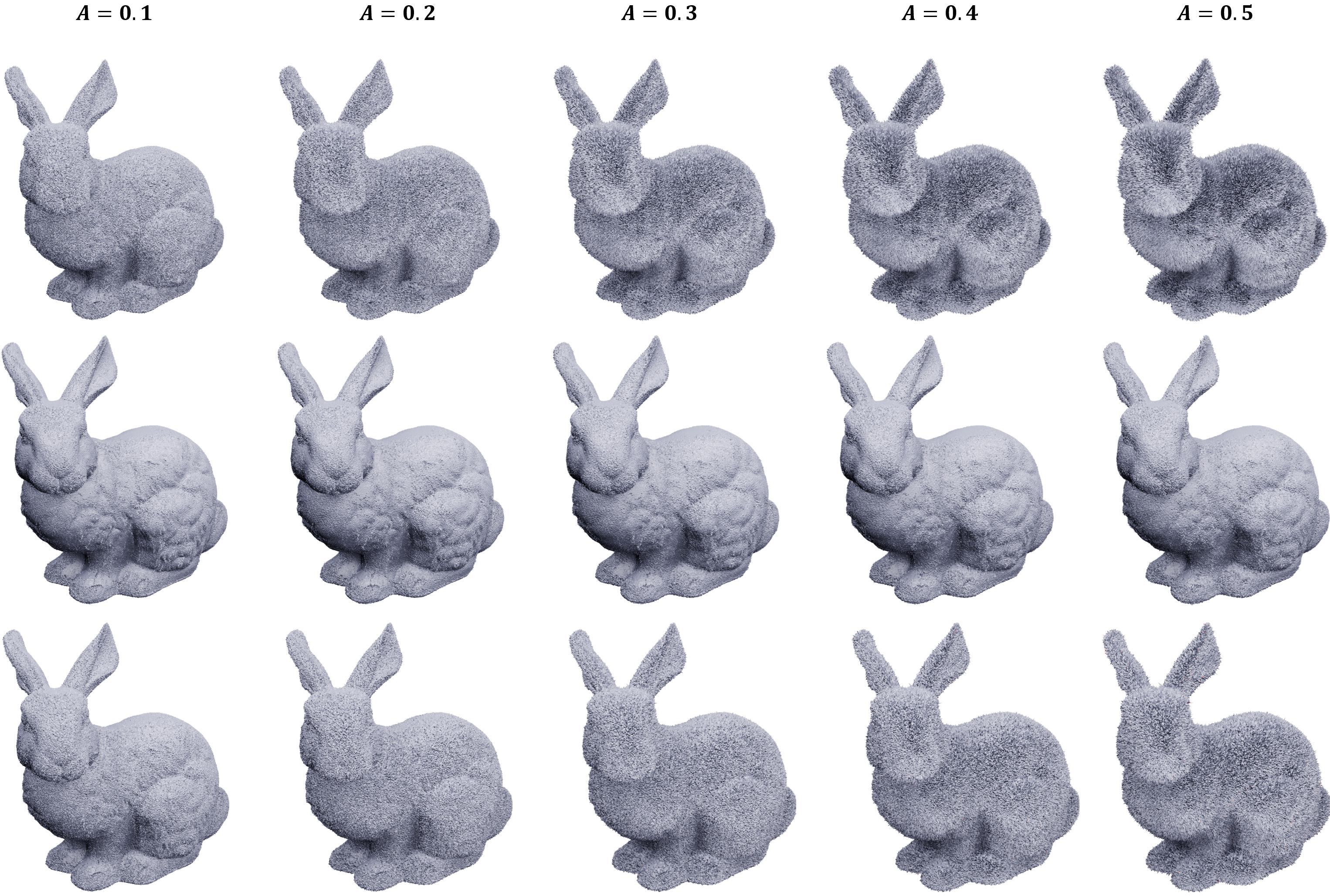}
    \caption{Noise experiment on the raw data of bunny. The backface is rendered in red. The first row shows the result of the experiment with noise affecting normal direction. The middle row depicts one in the tangential direction. While the last row is for fully random noise. The holes are rendered in red faces, it is hard to spot any holes with a glance. With careful inspection, only meshes with $A=0.5$ vertical and fully random noise have small failures.}
    \label{fig:position_noise}
\end{figure*}

\revision{In addition to applying random noise to all three coordinates of the point positions, we also did two experiments where noise was added separately in the tangential and normal directions. The results are shown in Figure~\ref{fig:position_noise} and seem to indicate that out method is quite resilient to noise applied to the positions.}

\subsubsection{Noise - normal direction}
\label{sec:normal-noise-test}
In this experiment, we apply noise solely to the normal direction. This involves selecting \revision{\sout{a random axis}the original normal direction as the axis of rotation} and pivoting the normal by a constant angle $\theta$ around this axis. Unsurprisingly, the outcome reveals that our method is more sensitive to variations in the normal direction. When pivoting the normal away by an angle of 15 degrees, significant holes appear in the results. 

This is relatively easy to explain. When a candidate edge is considered, the topology test checks if the edge belongs to a corner of the same face at either endpoint. Changing the normal confounds the rotation systems, and this means that even if a vertex belongs to the same face (as if the normals had been unperturbed) the candidate edge may not be incident on the vertex at a corner that belongs to the face.

In practice, though, this will never be an issue. The noise we applied was independent for each vertex, and this is the worst-case scenario for our reconstruction method since it relies on smoothly changing normals. Moreover, in practice, any method for normal computation would produce much smoother normal fields.


\begin{figure*}[h]
    \centering
    \includegraphics[width=\textwidth]{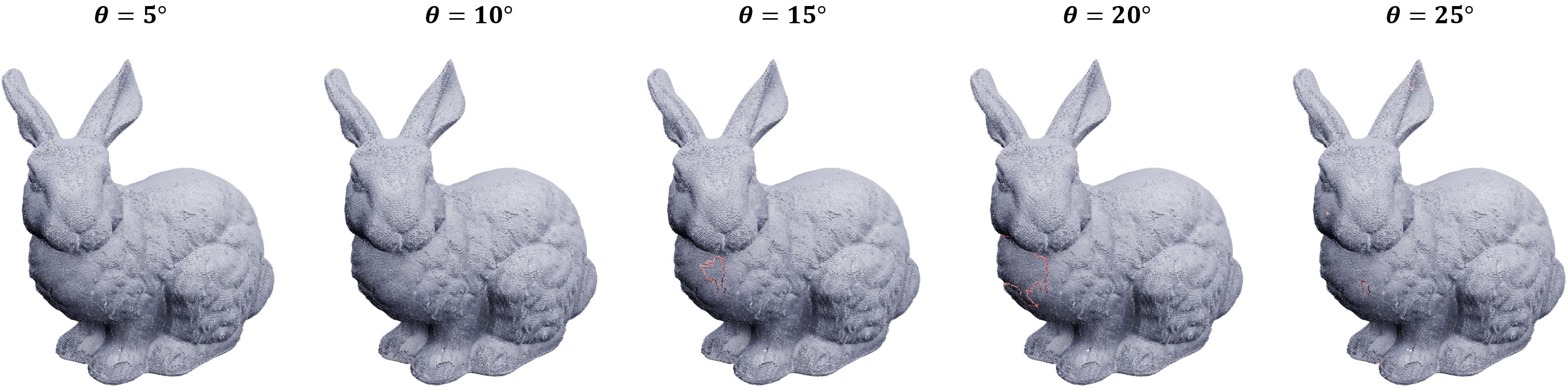}
    \caption{Noise experiment on normal direction. \revision{The normal vectors of the vertices are randomly rotated within an angle $\theta$ relative to their initial normal directions. As the noise becomes stronger, more connections are rejected, resulting in holes that are rendered in red.}}
    \label{fig:angle_noise}
\end{figure*}

\begin{figure*}[h]
    \centering
    \includegraphics[width=\textwidth]{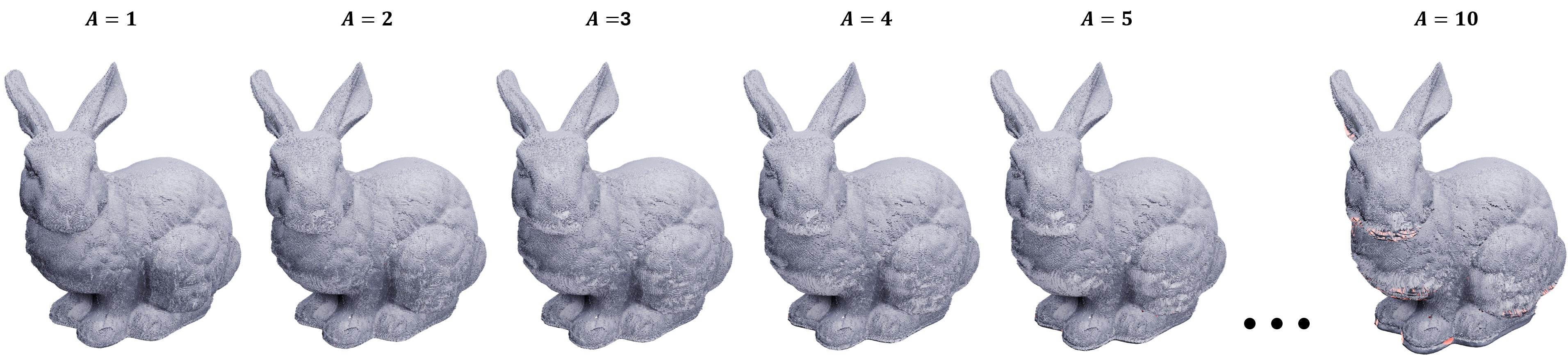}
    \caption{Misalignment (displacement) experiment on the raw data of Stanford bunny. \revision{$A$ represents the magnitude of the displacement. Another layer gradually shows up as $A$ increases.}}
    \label{fig:mis_a}
\end{figure*}


\subsubsection{Misalignment}
\paragraph{Displacement}
Another common defect in reconstruction is misalignment. To create artificial data, we take one point cloud file from the raw data of bunny (specifically, \textit{chin.ply}) as the misaligned part. The point cloud is then gradually moved away from the registered position. The displacement is defined as:
\begin{equation} 
    \mathbf{d} = A \cdot \overline{|e|} \cdot \mathbf{v} \enspace ,
\end{equation}
where $A$ is the changing amplitude, $\overline{|e|}$ is the average edge length, and $\mathbf{v}$ is the offset direction. Figure~\ref{fig:mis_a} presents the results. Our method effectively addresses misalignment when it is minimal (less than 4 times the average length). However, as misalignment increases, normal estimation is impacted, leading to the appearance of holes. When the misalignment reaches a distance 10 times the average edge length away from the original position, it is deemed to represent another layer of surface, resulting in failure.

\paragraph{rotation \& non-rigid misalignment}

\begin{figure*}[h]
    \centering
    \includegraphics[width=\textwidth]{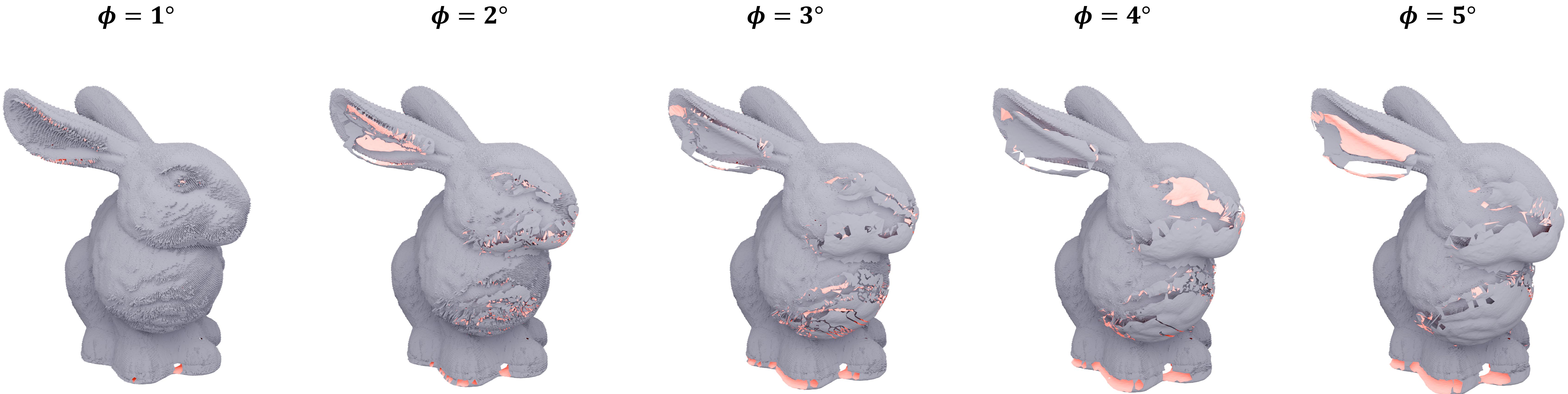}
    \caption{\revision{Misalignment (rotation) experiment on the raw data of bunny. As $\phi$ increases, two layers around the chin are reconstructed with holes and undesirable connections.}}
    \label{fig:mar}
\end{figure*}

Besides displacement, other types of misalignment exist including rotational and non-rigid misalignment. Both types will create a situation where multiple layers of vertices with different normals coinhabit the same local region. If the normals are estimated from the combined point cloud, it is impossible to avoid connections between points from different layers. This leads to a situation where vertices from different layers connect with each other, often leading to holes in the reconstruction since the layers are prone to interpenetration which, of course, is a non-manifold configuration. \revision{Figure~\ref{fig:mar} illustrates a similar experiment on Stanford bunny. Again, \textit{chin.ply} is rotated $\phi$ degrees around the z axis.}

\begin{figure}[h]
  \centering
  \includegraphics[width=1.0\columnwidth]{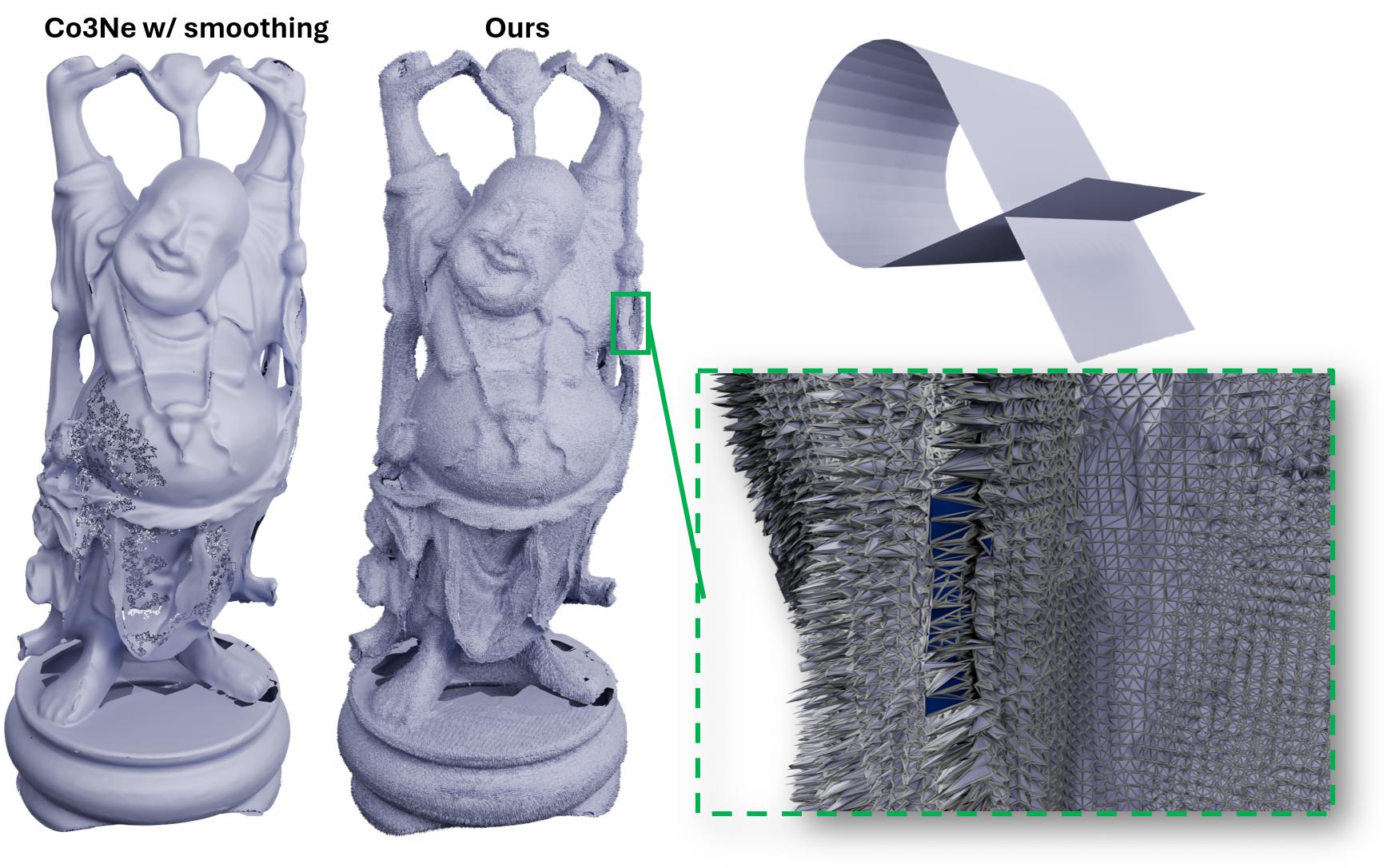}
  \caption{Left and bottom: reconstruction of the so-called Happy Buddha model from the original scan data is a fail case due to sub-scans that interpenetrate but still connect since points from all scans contribute to normal estimation. We provide the reconstruction result by Co3Ne w/ smoothing to show how challenging to reconstruct it. Right: synthetic case where a self-intersecting surface is resolved correctly by our approach by using exact normals to disambiguate.}
  \label{fig:intersection_case}
\end{figure}

\revision{A non-synthetic example is} the failed reconstruction from the raw data of Stanford Happy Buddha \revision{shown} in Figure~\ref{fig:intersection_case} (left). \revision{We note that if the normals had been correctly esitmated for each part and sufficiently different) our method would simply have meshed the different sub-scans and allowed them to intersect. A synthetic example illustrating this point is shown in Figure~\ref{fig:intersection_case} (right)}. However, often the layers should be combined, and then we get a much more stable normal estimate if it is based on the combined point cloud. Hence, it is generally better to estimate the normals on the combined point cloud.

\section{Discussion}
%
\revision{Our work is based on the simple observation that a tree connecting all points in a point cloud is a genus 0 graph and, given a rotation system, we can construct a mesh with a single face that contains all points}. The edges of that face are precisely the edges of the tree covered twice, and our results demonstrate that it is possible to triangulate this face by incrementally inserting edges. We can raise the genus by explicitly and judiciously adding handles.

The proposed method outputs a surface reconstruction where nearly all points that belong to the graph that we construct initially are also part of the output mesh. Many other methods reject a large number of points in order to construct the best possible output mesh, but in effect this means that reconstruction and noise reduction are subtly coupled. If this is desired it seems that volumetric methods might be more appropriate. In contrast, our method only rejects very few points and leaves it to the downstream processing to reduce noise. That can be beneficial for instance in applications where several sub-scans are combined, and it is desirable that there is no bias in how many point are kept from each sub-scan.

Compared to previous work, we initially make sure that connected vertices belong to the same face. This allows us to guarantee that gratuitous handles are not added. Of course, the true genus of the reconstructed object is not necessarily known. Hence, it is not generally possible to guarantee correct topology, but since we add handles explicitly, we can impose a threshold on the handle quality.

The fact that handles are added explicitly also makes it trivial to restrict the topology of reconstructed objects that are known to be of sphere or disk topology since we can simply omit the insertion of handles. The utility of this was demonstrated in Section~\ref{sec:topology-control} where we used topology control to restrict the cortical surface reconstruction to genus 0. For objects of higher genus, we do need to insert handles, but we can still restrict the topology by only inserting handles (in prioritized order) till the desired genus is attained.

Like many other methods, we are reliant on consistently oriented normals.  In principle, one could find a rotation system without normals, e.g. by searching for consistent circular orderings of edges at each vertex, but then the resulting rotation system would \textit{induce} a consistent normal orientation. In return for our dependence on normals, we get the benefit of being able to utilize surface normals to disambiguate surfaces which are geometrically very close such as the two sides of the terrain in Section~\ref{sec:synthetic_comp}.

\subsection{Limitations and Future Work}
Our method is relatively robust to noise on the positions of the points, but susceptible to normal noise. Fortunately, the normal field only has to be smooth and not necessarily accurate with respect to the underlying surface, and in many practical scanning scenarios where points are noisy, we can obtain smooth normal estimates by using a large number of nearby points.

The main failure mode of our method is when point clouds consist of several scans that are so poorly aligned that they interpenetrate.
In these cases, the underlying data is manifestly non-manifold, and, since we can only reconstruct manifold surfaces, this leads to faces that cannot be triangulated and, in turn, components of the mesh that disconnect from the rest of the model. In principle this could be remedied by only connecting points in the initial graph if either the points come from the same scan or their normals are very similar. This is something that we will consider more in future work.

A strong tool for furthering the practical performance of algorithms is that of parallelization. The current implementation is sequential, and while the algorithm isn't embarrassingly parallel, we see ample opportunities to make use of parallelization.

The main hindrances to parallelization lie in initialization, and the splitting of faceloops. Luckily, libraries like the Boost Graph Library provides implementations for parallel computations of minimum spanning trees, leaving us to deal with parallel maintenance of faceloops. Recall that faceloops are represented as balanced binary trees, that once split, will never interact. Therefore it suffices to lock access to the single faceloop that is being processed, leaving the remaining trees open for manipulation in parallel. This will cause some congestion initially, but as the number of faces grows, so does the potential for parallelization.

%
\begin{acks}
We thank all data providers: the Stanford Computer Graphics Laboratory; Qingnan Zhou and Alec Jacobson; Chu et al.; Armeni et al.; Matterport, Inc.; Knapitsch et al.; and Anderson Winkler. We thank the anonymous reviewers for their helpful comments. This work is partially supported by a DTU alliance scholarship, the Danish Council for Independent Research (6111-00552B), the Carlsberg Foundation (CF21-0302), and the German Research Foundation (Gottfried Wilhelm Leibniz programme).
\end{acks}

\bibliographystyle{ACM-Reference-Format}
\bibliography{bibliography}



\end{document}